\input harvmac
\input epsf.tex
\def\N{{\cal N}}
\def\O{{\cal O}}

\lref\GW{D.~J.~Gross and F.~Wilczek,
``Asymptotically Free Gauge Theories. 2,''
Phys.\ Rev.\ D {\bf 9}, 980 (1974).
}

\lref\BZ{T.~Banks and A.~Zaks,
``On The Phase Structure Of Vector - Like Gauge Theories With Massless
Fermions,''
Nucl.\ Phys.\ B {\bf 196}, 189 (1982).
}

\lref\AppelquistRB{
T.~Appelquist, A.~Ratnaweera, J.~Terning and L.~C.~Wijewardhana,
``The phase structure of an SU(N) gauge theory with N(f) flavors,''
Phys.\ Rev.\ D {\bf 58}, 105017 (1998)
[arXiv:hep-ph/9806472].
}

\lref\LeighEP{ R.~G.~Leigh and M.~J.~Strassler, ``Exactly marginal
operators and duality in four-dimensional N=1 supersymmetric gauge
theory,'' Nucl.\ Phys.\ B {\bf 447}, 95 (1995) [arXiv:hep-th/9503121].
}
\lref\LeighEP{
R.~G.~Leigh and M.~J.~Strassler,
``Exactly marginal operators and duality in four-dimensional N=1
supersymmetric gauge theory,''
Nucl.\ Phys.\ B {\bf 447}, 95 (1995)
[arXiv:hep-th/9503121].
}
\lref\KPS{
D.~Kutasov, A.~Parnachev and D.~A.~Sahakyan,
``Central charges and $U(1)_R$ symmetries in $\N = 1$ super
Yang-Mills,''
arXiv:hep-th/0308071.
}

\lref\ArgyresJJ{
P.~C.~Argyres and M.~R.~Douglas,
``New phenomena in SU(3) supersymmetric gauge theory,''
Nucl.\ Phys.\ B {\bf 448}, 93 (1995)
[arXiv:hep-th/9505062].
}
\lref\IntriligatorID{
K.~A.~Intriligator and N.~Seiberg,
``Duality, monopoles, dyons, confinement and oblique confinement in
supersymmetric SO(N(c)) gauge theories,''
Nucl.\ Phys.\ B {\bf 444}, 125 (1995)
[arXiv:hep-th/9503179].
}
\lref\IntriligatorER{
K.~A.~Intriligator and N.~Seiberg,
``Phases of N = 1 supersymmetric gauge theories and electric-magnetic
triality,''
arXiv:hep-th/9506084.
}

\lref\IW{
K.~Intriligator and B.~Wecht,
``The exact superconformal R-symmetry maximizes $a$,''
arXiv:hep-th/0304128.
}

\lref\JB{
J.~H.~Brodie,
``Duality in supersymmetric $SU(N_c)$ gauge theory with two adjoint
chiral
superfields,''
Nucl.\ Phys.\ B {\bf 478}, 123 (1996)
[arXiv:hep-th/9605232].
}

\lref\Cardy{
J.~L.~Cardy,
``Is There A C Theorem In Four-Dimensions?,''
Phys.\ Lett.\ B {\bf 215}, 749 (1988).
}

\lref\AEFJ{D.~Anselmi, J.~Erlich, D.~Z.~Freedman and A.~A.~Johansen,
``Positivity constraints on anomalies in supersymmetric gauge
theories,''
Phys.\ Rev.\ D {\bf 57}, 7570 (1998)
[arXiv:hep-th/9711035].
}

\lref\AFGJ{D.~Anselmi, D.~Z.~Freedman, M.~T.~Grisaru and A.~A.~Johansen,
``Nonperturbative formulas for central functions of supersymmetric gauge
theories,''
Nucl.\ Phys.\ B {\bf 526}, 543 (1998)
[arXiv:hep-th/9708042].
}

\lref\Zam{A.~B.~Zamolodchikov,
``'Irreversibility' Of The Flux Of The Renormalization Group In A 2-D
Field Theory,''
JETP Lett.\  {\bf 43}, 730 (1986)
[Pisma Zh.\ Eksp.\ Teor.\ Fiz.\  {\bf 43}, 565 (1986)].
}

\lref\NSVZ{V.~A.~Novikov, M.~A.~Shifman, A.~I.~Vainshtein and
V.~I.~Zakharov,
``Exact Gell-Mann-Low Function Of Supersymmetric Yang-Mills Theories
From
Instanton Calculus,''
Nucl.\ Phys.\ B {\bf 229}, 381 (1983).
}

\lref\HO{H.~Osborn,
``N = 1 superconformal symmetry in four-dimensional quantum field
theory,''
Annals Phys.\  {\bf 272}, 243 (1999)
[arXiv:hep-th/9808041].
}

\lref\GW{D.~J.~Gross and F.~Wilczek,
``Asymptotically Free Gauge Theories. 2,''
Phys.\ Rev.\ D {\bf 9}, 980 (1974).
}

\lref\BZ{T.~Banks and A.~Zaks,
``On The Phase Structure Of Vector - Like Gauge Theories With Massless
Fermions,''
Nucl.\ Phys.\ B {\bf 196}, 189 (1982).
}

\lref\NSd{N.~Seiberg,
``Electric - magnetic duality in supersymmetric nonAbelian
gauge theories,''Nucl.\ Phys.\ B {\bf 435}, 129
(1995)[arXiv:hep-th/9411149].
}

\lref\CKV{\
F.~Cachazo, S.~Katz and C.~Vafa,
``Geometric transitions and N = 1 quiver theories,''
arXiv:hep-th/0108120.
}

\lref\JE{J.~Erdmenger,
``Gravitational axial anomaly for four dimensional conformal field
theories,''
Nucl.\ Phys.\ B {\bf 562}, 315 (1999)
[arXiv:hep-th/9905176].
}

\lref\KINS{K.~A.~Intriligator and N.~Seiberg,
``Phases of N=1 supersymmetric gauge theories in four-dimensions,''
Nucl.\ Phys.\ B {\bf 431}, 551 (1994)
[arXiv:hep-th/9408155].
}

\lref\KINS{K.~A.~Intriligator and N.~Seiberg,
``Phases of N=1 supersymmetric gauge theories in four-dimensions,''
Nucl.\ Phys.\ B {\bf 431}, 551 (1994)
[arXiv:hep-th/9408155].
}

\lref\DKi{
D.~Kutasov,
``A Comment on duality in N=1 supersymmetric nonAbelian gauge
theories,''
Phys.\ Lett.\ B {\bf 351}, 230 (1995)
[arXiv:hep-th/9503086].
}

\lref\DKAS{
D.~Kutasov and A.~Schwimmer,
``On duality in supersymmetric Yang-Mills theory,''
Phys.\ Lett.\ B {\bf 354}, 315 (1995)
[arXiv:hep-th/9505004].
}

\lref\arnold{V.I. Arnold, S.M. Gusein-Zade and
A.N. Varchenko, {\it Singularities of Differentiable Maps} volumes
1 and 2, Birkhauser (1985), and references therein.}

\lref\DKNSAS{D.~Kutasov, A.~Schwimmer and N.~Seiberg,
``Chiral Rings, Singularity Theory and Electric-Magnetic Duality,''
Nucl.\ Phys.\ B {\bf 459}, 455 (1996)
[arXiv:hep-th/9510222].
}

\lref\thermalc{
T.~Appelquist, A.~G.~Cohen and M.~Schmaltz,
``A new constraint on strongly coupled field theories,''
Phys.\ Rev.\ D {\bf 60}, 045003 (1999)
[arXiv:hep-th/9901109].
}

\lref\superspace{S.~J.~Gates, M.~T.~Grisaru, M.~Rocek and W.~Siegel,
``Superspace, Or One Thousand And One Lessons In Supersymmetry,''
Front.\ Phys.\  {\bf 58}, 1 (1983)
[arXiv:hep-th/0108200].
}

\lref\ISrev{
K.~A.~Intriligator and N.~Seiberg,
``Lectures on supersymmetric gauge theories and electric-magnetic
duality,''
Nucl.\ Phys.\ Proc.\ Suppl.\  {\bf 45BC}, 1 (1996)
[arXiv:hep-th/9509066].
}
\lref\anselmi{D.~Anselmi,
``Inequalities for trace anomalies, length of the RG flow, distance
between
the fixed points and irreversibility,''arXiv:hep-th/0210124.
}

\def\ev#1{\langle#1\rangle}
\

\def\drawbox#1#2{\hrule height#2pt
             \hbox{\vrule width#2pt height#1pt \kern#1pt \vrule 
width#2pt}
                   \hrule height#2pt}

\def\Asym#1#2{\vcenter{\vbox{\drawbox{#1}{#2}
                   \kern-#2pt       
                   \drawbox{#1}{#2}}}}

\Title{\vbox{\baselineskip12pt\hbox{hep-th/0309201}
\hbox{UCSD-PTH-03-12} }}
{\vbox{\centerline{RG Fixed Points and Flows in SQCD with Adjoints
}}}
\centerline{ Ken Intriligator
and Brian Wecht}
\bigskip
\centerline{Department of Physics} \centerline{University of
California, San Diego} \centerline{La Jolla, CA 92093-0354, USA}

\bigskip
\noindent
We map out and explore the zoo of possible 4d $\N=1$ superconformal
theories which are obtained as RG fixed points of $\N=1$ SQCD with
$N_f$ fundamental and $N_a$ adjoint matter representations.  Using
``a-maximization," we obtain exact operator dimensions at all RG
fixed points and classify all relevant, Landau-Ginzburg type,
adjoint superpotential deformations.  Such deformations can be used to
RG
flow to new SCFTs, which are then similarly analyzed.
Remarkably,  the resulting 4d SCFT classification coincides
with Arnold's ADE singularity classification.
The exact superconformal R-charge and the central charge $a$ are
computed for all of these theories.  RG flows between the different
fixed points are analyzed, and all flows are verified to be compatible
with the conjectured $a$-theorem.

\Date{September 2003}
\newsec{Introduction}

Asymptotically free gauge theories have a variety of interesting
possible
IR phases\foot{Non-asymptotically free theories, on the
other hand, flow to free field theories in the IR (at least when one
starts the RG flow at weak coupling).}.
For example, they can flow to theories with a mass gap for the
gauge fields and confinement,
or to theories with no mass gap and an interacting
``non-Abelian Coulomb phase,'' i.e.  an interacting 4d CFT.
Perturbative analysis suggested the possibility of the latter
phase long ago, via the apparent weakly coupled RG fixed
points in theories designed to be just barely asymptotically free
\refs{\GW, \BZ}.  There might have then been some lingering doubts about
about whether or not interacting 4d CFTs really existed (theoretically)
at the non-perturbative level.  But, over the past decade, the study
of supersymmetric gauge theories (see e.g. \ISrev) has provided strong
evidence for the
existence of a zoo of non-perturbatively exact renormalization group
fixed points.

Not only do interacting RG fixed points exist, but they are in some
sense generic: general gauge theories with enough matter fields (but
not too many so as to spoil asymptotic freedom) are believed to flow
to interacting CFTs in the IR. A well-known example of this is
ordinary (non-supersymmetric) $SU(N_c)$ gauge theory with $N_f$
fundamental flavors, which flows to an interacting CFT in the range
$N_f^{min}<N_f<{11\over 2}N_c$, where it has been estimated that
$N_f^{min}\approx 4N_c$ \AppelquistRB.  Here the conformal range is
somewhat narrow, but it becomes wider when other representations are
included.  We will here focus on the case of supersymmetric theories,
where there are some powerful tools available.

The vast collection of possible interacting 4d $\N=1$ SCFTs and RG
flows among them remains relatively unexplored.  One well-studied
case is SQCD, which flows to interacting RG fixed points when the
number $N_f$ of fundamental flavors is in the range ${3\over
2}N_c<N_f<3N_c$ \NSd.  A generalization of this is to consider theories
with other matter representations; these theories provide interesting
testing grounds for exploring ideas in quantum field theory.  One idea
which will be of particular interest for the present paper is the
conjecture that there is a 4d analog of Zamolodchikov's
2d c-theorem \Zam: that there exists a ``central charge,'' which
counts the number of degrees of freedom of a quantum field theory and
monotonically decreases along RG flows to the IR, as degrees of
freedom are integrated out.  It is further conjectured \Cardy, with much
supporting evidence e.g. \refs{\AFGJ, \AEFJ, \IW, \KPS}, that an
appropriate such central charge (at least at RG fixed points) is the
coefficient\foot{Unfortunately
the name $``a"$ is standard for this term.  The name $``c"$ reserved for
a different curvature-square term, which is known to {\it not} obey the
4d version
of the c-theorem \refs{\AFGJ, \AEFJ}.}
$``a"$ of a certain curvature-squared term (the Euler density) of
the conformal anomaly on a curved space-time background\foot{
There is another quantity which has been
conjectured to always decrease in the IR: the thermal c-function of
\thermalc.  We will not dicuss this latter proposal here, because
supersymmetry does not provide a known way to
exactly compute the thermal c-function at
interacting RG fixed points (there is a known general expression only
for free field theories).}.
The {\bf conjectured} a-theorem is then that all RG flows satisfy
$a_{IR}<a_{UV}$.  We will often refer to ``a-theorem predictions," but
please keep in mind that there is presently no generally accepted
proof of the a-theorem -- it is a conjecture.

The superconformal algebra implies that $a$ can be exactly computed
simply in terms of the 't Hooft anomalies for the superconformal
R-symmetry, $a=3\Tr R^3-\Tr R$ \AEFJ.  We will here rescale our
definition of $a$ to eliminate a conventional factor of $3/32$.
Because 't Hooft anomalies can be exactly computed in terms of the
original UV spectrum ('t Hooft anomaly matching), $a$ can be exactly
determined provided that the superconformal R-symmetry is exactly
determined.  Determining the exact superconformal R-symmetry had been
a stumbling block in analyzing theories with more than one type of
matter representation.  In \IW\ we recently showed that the
superconformal $U(1)_R$ symmetry is simply determined by
``a-maximization'': letting $R_t$ be the most general trial R-symmetry
(incorporating mixing with all global flavor symmetries), the
superconformal $U(1)_R\subset SU(2,2|1)$ is the unique choice of $R_t$
which locally maximizes
\eqn\atrial{a_{trial}(R_t) = 3 \Tr R_t^3 - \Tr R_t.}
    The value of $a_{trial}$ at the local maximum is then the central
charge $a$.  Using this result, it is now possible to apply the
powerful constraints of the superconformal algebra to explore these
more general 4d $\N=1$ SCFTs.  The result also almost proves the
a-theorem for SCFTS, up to some caveats \IW.  For a brief review of the
a-maximization
procedure, see Appendix B.

In \KPS, Kutasov, Parnachev, and Sahakyan applied and extended the
method of \IW\ to study $\N =1$ SQCD with $N_f$ fundamentals, $Q_i$
and $\widetilde Q_i$, and $N_a=1$ additional adjoint chiral
superfield $X$. Our work here was inspired by their detailed analysis
of the various possible RG fixed points and flows, and the many
striking ways in which the conjectured a-theorem was found to be
indeed always satisfied: for every RG flow, $a_{IR}<a_{UV}$.  An
important part of the analysis of \KPS\ was determining when gauge
invariant chiral primary operators $M$ appear to violate the unitarity
bound $R(M)={2\over 3}\Delta (M)\geq {2\over 3}$.  The belief
\refs{\NSd, \DKi, \DKAS, \DKNSAS} is that such operators are actually
free fields, with $R(M)$ corrected to be $2/3$ by an accidental $U(1)_M$
symmetry under which only the free field $M$ is charged.  It was
shown in \KPS\ that the effect of any such accidental symmetries must
be included in the a-maximization method, and that they non-trivially
affect the value for the superconformal R-charges and the central
charge $a$.  This will be important in some of our examples.

In this paper, we study the larger family of RG fixed points and flows
which can be obtained from $SU(N_c)$ SQCD with $N_f$ fundamentals
together with $N_a=2$ adjoint matter chiral superfields, $X$ and $Y$.
$N_a=2$ is the maximum number of adjoints compatible with asymptotic
freedom\foot{Of course, there are also RG fixed points for $N_a=3$
adjoints and $N_f=0$ flavors, provided that one adds the cubic
superpotential of the $\N =4$ theory or generalizations \LeighEP.}, so
this is a fairly complete study of the full family of possible RG
fixed points and flows which can be obtained via SQCD with fundamental
and adjoint representations.

The possible RG fixed points which we find and analyze can be
summarized as follows: \eqn\oade{\matrix{&\widehat O\cr &\widehat A\cr
&\widehat D\cr &\widehat E\cr &A_k\cr &D_{k+2}\cr
&E_6\cr &E_7\cr & E_8}\qquad \matrix{
&W_{\widehat O}
=0\cr &W_{\widehat A}=\Tr Y^2\cr &W_{\widehat
D}=\Tr XY^2\cr &W_{\widehat E}=\Tr Y^3\cr &W_{A_k}=\Tr
(X^{k+1}+Y^2) \cr &W_{D_{k+2}}=\Tr (X^{k+1}+XY^2) \cr &W_{E_6}
=\Tr (Y^3+X^4) \cr &W_{E_7}=\Tr (Y^3+YX^3) \cr &W_{E_8}
=\Tr (Y^3+X^5).\cr}}
There are additional RG fixed points associated with adding
Yukawa-type superpotentials involving the quarks; we will
briefly mention some of these possibilities in a later section.
The names given to
the fixed points in \oade\ are motivated by Arnold's \arnold\ $ADE$
classification of singularities, which precisely coincides with the
possible relevant deformation superpotentials, listed as $A_k$,
$D_{k+2}$
and $E_k$ in  \oade.  Our method for obtaining
the above superpotentials was based on a detailed analysis of the
anomalous
dimensions of operators at each of the RG fixed points, and when the
associated Landau-Ginzburg superpotentials
can be relevant and drive the theory to a new RG fixed point.  On the
face of it,
this has no obvious connection to any of the other known ways in which
Arnold's singularities
have appeared in mathematics or physics.

The possible RG flows between the above fixed points
are as shown in Figure 1. In this terminology, the work \KPS\ studied
the $\widehat A$ fixed points and RG flows to the $A_k$ fixed points,
as well as RG flows from $A_k$ fixed points to $A_{k'}$ fixed
points with $k'<k$.  We will extend this analysis to consider all of
the RG fixed points and flows of Figure 1.

\bigskip
\centerline{\epsfxsize=0.20\hsize\epsfbox{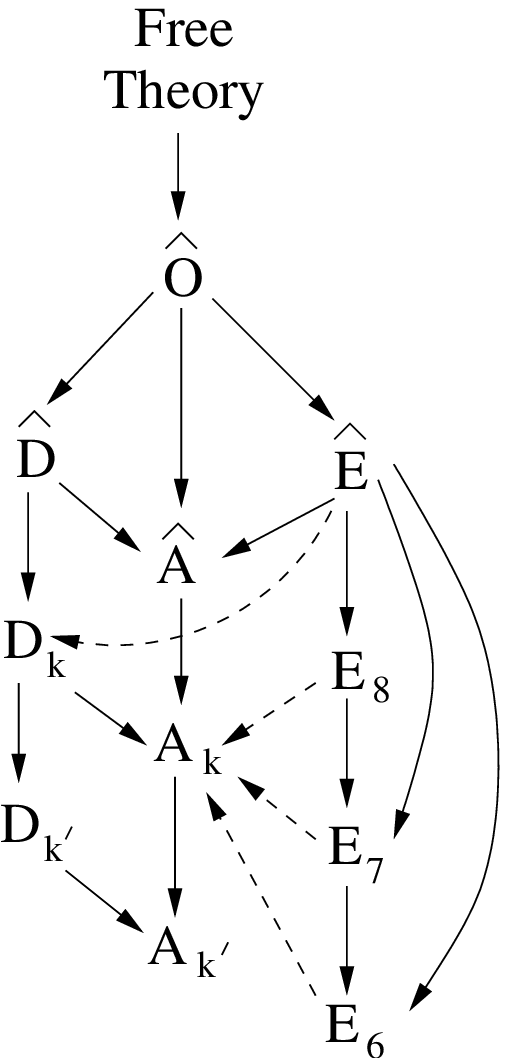}}
\centerline{\ninepoint\sl \baselineskip=8pt {\bf Figure 1:}
{\sl The map of possible flows between fixed points.}}
\centerline{\ninepoint\sl Dotted lines indicate flow to a particular
value
of $k$.}
\bigskip

The outline of this paper is as follows.  In sect. 2 we discuss
$SU(N_c)$ gauge theory with $N_f$ fundamental flavors, two adjoints,
and $W_{tree}=0$.  For all $N_f$ in the asymptotically free range,
$0\leq N_f<N_c$, these theories flow to interacting RG fixed points in
the IR.  We refer to these 4d ${\cal N}=1$ SCFTs as $\widehat O (N_c,
N_f)$,
or simply $\widehat O$.  We obtain the exact superconformal $U(1)_R$
charge and the exact central charge $a_{\widehat O}(x)$ in the limit
$N_c\gg 1$, for arbitrary fixed $x\equiv N_c/N_f$. None of our methods
actually depends on large $N_c$; it is merely employed to simplify the
exact expressions.  As $x$ increases, the RG fixed point is at
stronger coupling.  As might have been expected from the negative
anomalous dimensions of gauge theories, we find that
the superconformal $U(1)_R$ charges are monotonically decreasing
functions of $x$.  For all $x$,
we find no apparent unitarity violations.  Therefore it is not
necessary for there to be any free-field operators or associated
accidental symmetries for the $\widehat O$ RG fixed points.  We verify
that several a-theorem predictions are indeed satisfied.  Finally, we
classify and discuss all of the possible relevant superpotential
deformations of the $\widehat O$ RG fixed point which involve only the
adjoint fields $X$ and $Y$.  The interesting possibilities are the
superpotentials labeled $\widehat A$, $\widehat D$, and $\widehat E$
in \oade.

In sect. 3 we discuss the $\widehat E$ SCFTs, which are the IR
endpoints of an RG flow obtained by perturbing the $\widehat O$ SCFT
by the  relevant superpotential
$W_{\widehat E}=\Tr Y^3$.  We
obtain the exact superconformal R-charge and central charge
$a_{\widehat E}(x)$ of the $\widehat E$ SCFTs.  There are no apparent
unitarity violations for all $x$, and thus there is no need for any
free-field operators or associated accidental symmetries.  We verify
the a-theorem prediction, $a_{\widehat O}(x)>a_{\widehat E}(x)$, for
all $x>1$.
We also classify the relevant superpotential deformations of the
$\widehat E$ RG fixed point which involve only the adjoint field.  The
possibilities are indicated in Fig. 1.  The $E_6$, $E_7$, and $E_8$
deformations are only relevant provided that $x$ is sufficiently
large: $x>x_{E_6}^{min}\approx 2.44$, $x>x_{E_7}^{min}\approx 4.12$,
and $x>x_{E_8}^{min}\approx 7.28$, respectively.

In sect. 4 we discuss the $\widehat D$ SCFTs, which are the IR
endpoints of a RG flow obtained by perturbing the $\widehat O$ SCFT by
   the relevant superpotential
$W_{\widehat D}=\Tr XY^2$.  Unlike the
$\widehat O$ and $\widehat E$ RG fixed points, here we do find
apparent unitarity violations, indicating that various mesons
necessarily become free fields as $x$ increases.  We account for the
effect of the associated accidental symmetries by the procedure of \KPS\
(reviewed in appendix B) to obtain the exact superconformal $U(1)_R$
charges and exact central charge $a_{\widehat D}(x)$.  We verify
(numerically) the a-theorem prediction that $a_{\widehat
O}(x)>a_{\widehat D}(x)$ for all $x\geq 1$.  We also classify and
discuss
the various relevant deformations of the $\widehat D$ RG fixed point.
A class of such deformations is  $\Delta W=\Tr X^{k+1}$ which,
for any $k$, is a relevant deformation of $\widehat D$ provided that
$x$ is sufficiently large:  $x>x_{D_{k+2}}^{min}$.  We discuss how to
compute the lower bounds
$x_{D_{k+2}}^{min}$ and find e.g. that
$x_{D_{k+2}}^{min}\rightarrow {9\over 8}k$ for $k\gg 1$.

In sect. 5 we discuss the $D_{k+2}$ SCFTs, which are the IR endpoints
of the RG flow starting from the $\widehat D$ SCFT in the UV, upon
perturbing $W_{\widehat D}$ by $\Delta W=\Tr X^{k+1}$.  The $D_{k+2}$
SCFT exists if $x>x_{D_{k+2}}^{min}$, when
$\Delta W$ is a relevant $\widehat D$ deformation.
The superconformal $U(1)_R$ charges at the $D_{k+2}$ fixed point
are determined by $W_{D_{k+2}}$, and it is thus seen that a variety of
mesons apparently violate the unitarity bound and hence must be free.
The central charge $a_{D_{k+2}}(x)$ must be corrected, as in \AEFJ, to
account for
the free mesons.  We numerically verify the a-theorem prediction,
$a_{\widehat D}(x)> a_{D_{k+2}}(x)$ for all $x>x_{D_{k+2}}^{min}$,
plotting the example of
$a_{\widehat D}(x)>a_{D_5}(x)$.   At the $D_{k+2}$ RG fixed point there
is
a relevant deformation by $\Delta W=\Tr X^{k'+1}$, for $k'<k$, which
leads to the RG flow $D_{k+2}\rightarrow D_{k'+2}$.
We discuss this flow in Section 5 and the a-theorem prediction that
$a_{D_{k+2}}(x)>a_{D_{k'+2}}(x)$ for all $k'<k$ and
$x>x^{min}_{D_{k+2}}$.
There are apparent violations
of the a-theorem, but we verify that they always occur for
$x<x_{D_{k+2}}^{min}$, which is outside
of the range of validity needed for the $D_{k+2}$ RG fixed point to
exist.

In sect. 6 we discuss a magnetic dual description of the $D_{k+2}$
SCFTs due to Brodie \JB.  We determine the exact superconformal
R-charges and central charge $\widetilde a_{D_{k+2}}(x)$ in the
magnetic dual.  Our analysis
sheds light on the meaning of this duality, and when the various terms
in the magnetic superpotential are or are not relevant.  We show that
for all $k$ there is a conformal window
$x^{min}_{D_{k+2}}<x<3k-\widetilde x^{min}_{D_{k+2}}$ where the
$D_{k+2}$ RG fixed point is fully interacting, with the full
$W_{D_{k+2}}$
present in both the electric and magnetic descriptions. For large $k$
this conformal window is approximately given by ${9\over
8}k<x<(2.062)k$.

In sect. 7 we discuss the $E_6$, $E_7$, and $E_8$ SCFTs, which arise
as the IR limits of relevant deformations of the $\widehat E$ SCFTs upon
perturbing the $\widehat{E}$ theory by the superpotentials
$\Delta W = \Tr X^4$, $\Delta W = \Tr YX^3$, and $\Delta W = \Tr X^5$,
respectively.
These SCFTs only exist if $x$ is sufficiently large:
$x>x_{E_6}^{min}\approx 2.44$, $x>x_{E_7}^{min}\approx 4.12$, and
$x>x_{E_8}^{min}\approx 7.28$, respectively, as mentioned above.
There are also RG flows between these theories: $E_8\rightarrow E_7$,
$E_8
\rightarrow E_6$, $E_7\rightarrow E_6$.  Thus the a-theorem
prediction is $a_{\widehat E}(x)>a_{E_8}(x)>a_{E_7}(x)>a_{E_6}(x)$, at
least for the range of $x$ where each RG fixed point exists.  We find
that there are apparent violations of some of these a-theorem
predictions for
some ranges of $x$, but these ranges are always outside of the range
of validity $x>x^{min}$ required for the UV SCFT to exist.

In sect. 8 we briefly discuss additional RG flows and fixed points
which can be
obtained from those of \oade\ by perturbing by Yukawa-type interactions.

In sect. 9 we discuss the possibility that our procedure which led to
\oade,
starting at the $\widehat O$ SCFT and branching out to new SCFTs, as in
fig. 1, could miss some additional SCFTs.  This could happen if these
hypothetical
new SCFTs do not have $\widehat O$ in their domain of attraction.  As a
concrete
example, we consider the theory with $W=\lambda \Tr Y^{k+1}$ for $k>2$.
   For
small $\lambda$, this superpotential is an irrelevant deformation of
$\widehat O$ and
does not lead to a new SCFT.  But perhaps there is nevertheless an RG
fixed point
for some large critical value of $\lambda _*$.  We find this to be
unlikely but,
anyway, discuss the properties that such hypothetical new SCFTs should
have,
in order that their RG flows down to our other SCFTs in \oade\ be
compatible
with the a-theorem conjecture.

In sect. 10 we make some closing comments.

Finally, we have several appendices.  In appendix A, we discuss
the $\widehat O$, $\widehat D$, $\widehat E$, and $D_4$ RG fixed points
in the perturbative regime, where the theory is just barely
asymptotically free: $x=1+\epsilon$, with $0<\epsilon\ll 1$. In
appendix B, we review a-maximization \IW\ and the necessary modification
when there are accidental symmetries associated with operators becoming
free fields \KPS.  In appendix C, we discuss baryon operators and
whether or not they ever potentially violate the unitarity bound
in our various SCFTs.  In appendix D, we discuss the magnetic
duals of the $D_{k+2}$ RG fixed points in an asymptotic regime in
order
to determine the upper bound for the $D_{k+2}$ interacting conformal
window: ${9\over 8}k<x<(2.062)k$ for large $k$.

\newsec{The $\widehat O$ RG fixed points: $W(X,Y)=0$.}

The theories with $W_{tree}=0$ are expected\foot{ An
argument \KINS\ for this is to deform the theory by $W=m\Tr Y^2+\lambda
\sum _i \Tr \widetilde Q_i X Q_i$, making the IR theory $\N =2$ SQCD.
That theory has a moduli space with massless
monopoles and dyons at various places.  Upon taking $m\rightarrow 0$
and $\lambda \rightarrow 0$, the massless monopole and dyon locations
all collapse to the origin of the moduli space.  The presence of
massless fields which are not mutually local at the origin is believed
to signify the presence of an interacting RG fixed point \KINS.} to flow
to an interacting RG fixed point, which we call $\widehat O$,
for all $N_f$ in the asymptotically free range: $0\leq N_f<N_c$.

To simplify the formulae of this paper, we take $N_c$ large, with
$x\equiv N_c/N_f$ fixed.  Large $N_c$ is not essential to any of the
methods, it just simplifies the expressions for the results. It would be
straightforward, though tedious, to work with arbitrary $N_c$.

In the limit where the theory is just barely asymptotically free, $x=1+
\epsilon$ with $0<\epsilon \ll 1$, the $\widehat O$ RG fixed point is
at weak coupling; $\beta (g_*)=0$ for $g_*^2N\ll 1$.  The $\widehat O$
SCFT can then be analyzed in perturbation theory, as will be done in
appendix A.  As we increase $x$, the $\widehat O$ RG
fixed point is at stronger and stronger coupling.
Powerful methods associated with supersymmetry will be used to
obtain exact results for all $x$.

In the extreme case of $N_f=0$, where $x\rightarrow \infty$,
the Lagrangian has a unique candidate
superconformal R-symmetry which is anomaly free and commutes with the
flavor symmetries: $R(X)= R(Y)=\half$.  The true
superconformal R-symmetry could potentially be modified by
accidental symmetries, though there are no unitarity bound violations
which would require this to happen.  For the case $N_c=2$, this RG
fixed point has known electric, magnetic, and dyonic dual descriptions
\refs{\IntriligatorID , \IntriligatorER}.  We might expect analogous
dual descriptions for higher $N_c$, corresponding to the various
$Z_N\times Z_N$ electric and magnetic center phases, but this is not
presently known.

\subsec{The chiral ring of operators}

The set of ``observables" of SCFTs is the spectrum of gauge invariant
operators and their correlation functions.
A special set of such operators for SCFTs are the chiral primary
operators,
whose dimension is related to their R-charge by $\Delta = {3\over 2}R$.

In our theories, the microscopic chiral fields are the two adjoints $X$
and $Y$, the quarks $Q_i$ in the fundamental and $\widetilde Q_{\tilde
i}$ in the anti-fundamental, and the chiral gauge field strength
$W_\alpha$.  The chiral ring of operators is the set of all
gauge invariant
composites of these fields.  For example, we can form
\eqn\oops{{\cal O}_{I_1\dots I_n}=\Tr X_{I_1}\dots X_{I_n},}
where $I_i=1,2$ labels the two adjoints, which we also call $X$ and
$Y$.  We can form arbitrary such products of operators, with
$n$ arbitrarily large, though there will be some relations among them
for $n$ of order $N_c$: the space of expectation values of these
operators, subject to their classical relations, parameterize the
$N_c^2-1$ dimensional classical moduli space of vacua
where $X$ and
$Y$ have D-flat expectation values, with $\ev{Q_i}=\ev{\widetilde
Q_i}=0,$
along which $SU(N_c)$ is completely Higgsed.

Another class of gauge invariant operators is the set of generalized
``mesons"
\eqn\mesgen{(M_{I_1\dots I_n})_{i\tilde i}=\widetilde Q_{\tilde
i}X_{I_1}
\dots X_{I_n}Q_i.}
Finally, there is another class of operators, the generalized baryons,
which can be formed from the various dressed quarks:
\eqn\dressq{(Q_{(I_1\dots I_n)})_i\equiv (X_{I_1}\dots X_{I_n}Q)_i.}
We make baryons by
contracting $N_c$ dressed quarks with the $SU(N_c)$ epsilon tensor:
\eqn\barygeni{\eqalign{B&=Q_{(I_1\dots I_{n_1})}^{n_{(I_1\dots
I_{n_1})}}
Q_{(J_1\dots
J_{n_2})}^{n_{(J_1\dots J_{n_2})}}\dots Q_{(K_1\dots K_{n_k})}^{n_{(K_1
\dots I_{n_k})}}, \cr
\quad \hbox{with}\quad N_c&=n_{(I_1\dots I_{n_1})}+\dots +n_{(K_1\dots
K_{n_k})}\qquad\hbox{and all}\quad n_{(I_1\dots I_{n_1})}\leq N_f .}}

\subsec{Finding the exact superconformal $U(1)_R$ by a-maximization}

The theory with $N_f=2$ fundamentals and $N_a=2$ adjoints has an
$SU(N_f)\times SU(N_f)\times U(1)_B\times SU(2)_a\times U(1)_X \times
U(1)_R$ global flavor symmetry.  As reviewed in Appendix B,
we find the superconformal $U(1)_R$ symmetry by maximizing the
combination of 't Hooft anomalies \atrial\
using a general trial R-symmetry. We'll
parameterize the general anomaly-free trial R-symmetry as
\eqn\rtwoadj{R(Q_i) = R(\widetilde{Q}_i) \equiv  y, \quad R(Y) \equiv
z, \quad
R(X) = 1 - z + { 1-y \over x},} with $x\equiv N_c/N_f$.
$a$-maximization always yields a superconformal R-symmetry which
commutes
with all non-Abelian flavor symmetries, as well as charge conjugation
\IW. Thus, we could have restricted our trial $U(1)_R$ \rtwoadj\ to
commute
with $SU(2)_a$, i.e. $R(X)=R(Y)$ and hence $z=\half (1+(1-y)/x)$, at the
outset.  We chose to not impose this
condition in \rtwoadj\ for later reference, when we deform by
superpotential terms which break $SU(2)_a$.

Computing $a_{trial}$ \atrial\ with the R-charges \rtwoadj\ yields
(taking
$N_c$ large, $x$ fixed)
\eqn\atwoadj{\eqalign{ {a \over N_f^2} &= 2x^2 +
3x^2(z-1)^3+3x^2\left ( {1-y \over x} - z \right )^3 -x^2(z-1) \cr &
\quad -x^2\left ( {1-y \over x} - z \right ) + 6x(y-1)^3-2x(y-1).}}
Maximizing this with respect to $y$ and $z$ yields
\eqn\maxy{R(Q)=R(\widetilde Q)\equiv y(x)= 1+{ 3x
-2x\sqrt{26x^2-1} \over 3(8x^2-1)}} and \eqn\maxz{R(X)=R(Y)\equiv
z(x)= {1 \over 2} \left ( 1 + {-3+2\sqrt{26x^2 -1} \over 3(
8x^2-1)} \right ).}  Plugging these R-charges back into the central
charge gives
\eqn\amaxi{
{a _{\widehat O}(x)\over N_f^2} = {2x^2
\left (18 + 648x^4 - 2\sqrt{26x^2 -1}+x^2(-279+52\sqrt{26x^2-1}) \right
)
\over 9(1-8x^2)^2}.}

Both $y(x)$ and $z(x)$ are strictly decreasing functions of $x$.  In
the limit $x \rightarrow \infty$,
\eqn\ohatrlim{y(x\rightarrow \infty)\rightarrow 1-{\sqrt{26}\over 12}
\approx 0.575, \qquad z(x\rightarrow \infty)\rightarrow \half.}
We have plotted these functions in fig. 2 and fig. 3, respectively.
\bigskip
\centerline{\epsfxsize=0.50\hsize\epsfbox{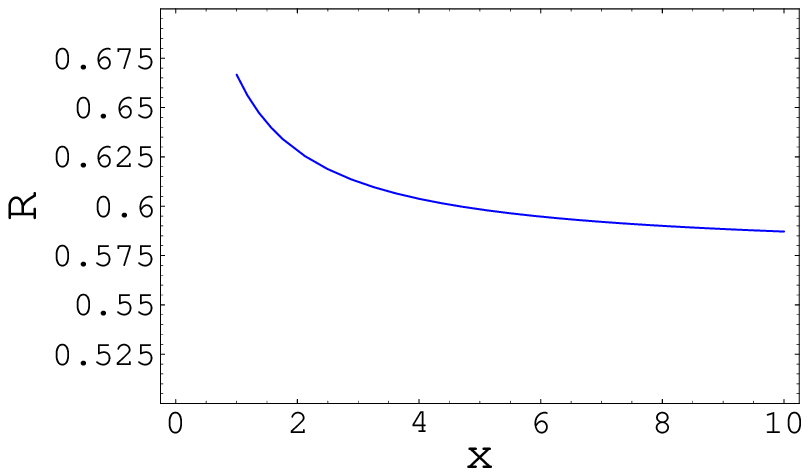}}
\centerline{\ninepoint\sl \baselineskip=8pt {\bf Figure 2:}
{\sl $y(x)=R(Q)$ for the $\widehat{O}$ theory.}}
\bigskip

\bigskip
\centerline{\epsfxsize=0.50\hsize\epsfbox{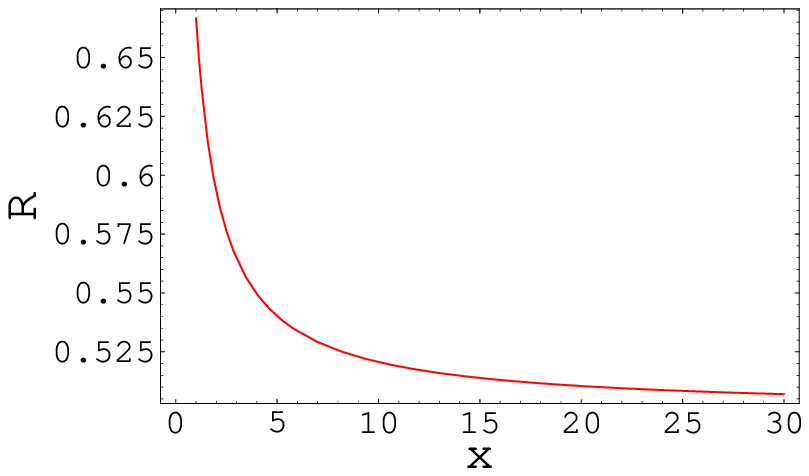}}
\centerline{\ninepoint\sl \baselineskip=8pt {\bf Figure 3:}
{\sl $z(x)=R(Y)=R(X)$ for the $\widehat{O}$ theory.}}
\bigskip
\noindent In
fig. 4 we plot $N_f^{-2}a_{\widehat O}(x)$, given by \amaxi, along
with that of the $g=0$ gauge coupling free field theory, where
$R(Q)=R(\widetilde Q)=R(X)=R(Y)={2\over 3}$:
\eqn\afreeuv{{a _{free}\over N_f^2} = {22 \over 9}x^2 + {4 \over 9}x.}
Because RG flow connects the free theory in the UV to the $\widehat O$
SCFT in the IR, the conjectured a-theorem prediction is $a_{\widehat
O}(x)<
a_{free}(x)$ for all $x\geq 1$; as seen from fig. 4, this is indeed
satisfied.
\bigskip
\centerline{\epsfxsize=0.50\hsize\epsfbox{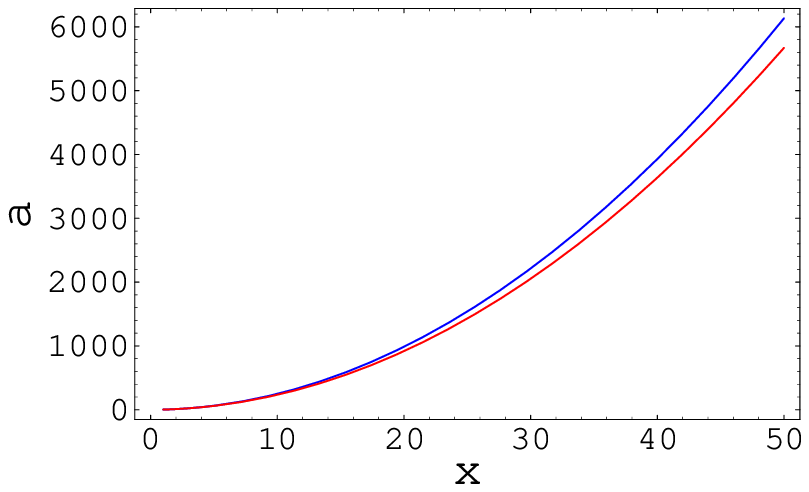}}
\centerline{\ninepoint\sl \baselineskip=8pt {\bf Figure 4:}
{\sl $a/N_f^2$ for the free UV (top, blue) and $\widehat{O}$ (bottom,
red) theories.}}
\bigskip

Because the $y(x)$ and $z(x)$ which we have found satisfy
$y(x)>\half$ and $z(x)>\half$ for all $x\geq 1$,
all gauge invariant chiral operators ${\cal O}$ in the chiral ring
satisfy the unitarity bound $R({\cal O})\geq 2/3$.
There is thus no reason to
doubt the validity of the above expressions for any $x$.  Had there
been apparent unitarity violations, we would have had to correct the
above
formulae to account for accidental symmetries, as in \KPS\ and reviewed
in
Appendix B.   It is still possible that there are actually some unseen
accidental symmetries for sufficiently large $x$ (large coupling) which
would modify the above results.  But we will see that our results are
consistent with various checks, such as the conjectured a-theorem, for
all $x\geq 1$, without any such modifications. So
we will tentatively propose that the above expressions
are exactly correct as given for all $1 <  x < \infty$.

One can easily generalize the above analysis to all $SU(N)$,
$SO(N)$, and $Sp(N)$ gauge theories, with $N_f$ fundamentals
and $N_a$ matter fields in a representation having quadratic index
of order $N$ (e.g. for $SU(N)$ we could replace each adjoint with a
$\bf{\half N(N\pm 1)+\overline{\half N(N\pm 1)}}$).  As long as $N_a=2$,
in all such cases the unitarity bound is found to be satisfied,
without any
need for accidental symmetries.

\subsec{Some checks of the conjectured $a$ theorem}

Let's do some more checks of the conjectured a-theorem,
$a_{UV}>a_{IR}$, in these examples.  Let
$a_{\widehat O}(N_c, N_f)$ be the central charge of the
$\widehat O$ RG fixed point for general $N_c$ and $N_f$,
which is given by \amaxi\ in our limit of large $N_c$ and $N_f$ with
$x\equiv
N_c/N_f$ fixed.

The a-theorem conjecture requires
\eqn\oapred{a_{\widehat O}(N_c, N_f)>a_{\widehat O}(N_c, N_f-1) \qquad
{\rm and} \qquad
a_{\widehat O}(N_c, N_f)>a_{\widehat O}(N_c-1, N_f+1).}
The first comes from the RG flow associated with giving a mass to one
of the fundamental quarks, $\Delta W=mQ_{N_f}\widetilde Q_{N_f}$, and
integrating it out.  The second prediction in \oapred\ comes from
Higgsing, going along a flat direction where we give a vev to
one of the fundamental flavors, or to some components of either adjoint.
When we Higgs $SU(N_c)$ to $SU(N_c-1)$, $N_f\rightarrow N_f-1+2$, with
one fundamental eaten, but two more coming from the two adjoints.

In the limit of large $N_c$ and $N_f$, we write the flow of the first
prediction in \oapred\ as
\eqn\flowi{N_f^{-2}\rightarrow N_f^{-2}(1+{2\over N_f}), \qquad
x\rightarrow
x(1+{1\over N_f}).}  Then the first case in \oapred\ can be written as:
\eqn\flowii{x^{-2}{a_{\widehat O}(x)\over N_f^2}\qquad
\hbox{must be a monotonically
{\it decreasing} function of $x$}.}
Likewise, the flow of the second prediction in \oapred\ is
\eqn\flowj{N_f^{-2}\rightarrow N_f^{-2}(1-{2\over N_f}), \qquad
x\rightarrow x(1-{2\over N_f}).}
The second prediction in \oapred\ is then
\eqn\flowjj{x^{-1}{a_{\widehat O}
(x)\over N_f^2}\qquad \hbox{must be a monotonically
{\it increasing} function of $x$}.}
Both of these predictions are indeed satisfied by our expression \amaxi\
for all $x \geq 1$.

Another a-theorem prediction is obtained by Higgsing $SU(N)\rightarrow
U(1)^{n-1}\times
\prod _{i=1}^n SU(N_i)$, for arbitrary $n$ and $N_i$ such that $\sum
_{i=1}^n
N_i=N$, by an $\ev{X}$ adjoint vev.  Each $SU(N_i)$ theory
has two adjoints and $N_f$ fundamentals, much as the original theory.
But,
in addition, there is a bi-fundamental flavor in the $({\bf N_i},
\overline
{\bf N_j})+ (\overline {\bf N_i}, {\bf N_j})$ for every pair of gauge
groups $SU(N_i)$ and $SU(N_j)$, which come from
decomposing $Y$ (the corresponding components of $X$ are eaten).
With so many bi-fundamentals, the $SU(N_i)$ theories generally are not
asymptotically free; $SU(N_i)$ is asymptotically free only if $
N_f + \sum_{j \neq i}N_j < N_i$,
which generally is not satisfied, e.g. if all $N_i$ are of the same
order.  When the $SU(N_i)$ are not asymptotically
free in the UV, they are free in the IR, and then $a_{IR}$ is simply
given in terms
of the free-field contributions:
\eqn\vevair{{a_{IR}(x)\over N_f^2} = {16 \over 9} \sum_i x_i^2 + {4
\over 9} x^2 + {4 \over 9}x,}
where $x_i \equiv N_i / N_f$.  As an example, we check the $a$-theorem
prediction,
that $a_{\widehat O}(x)>a_{IR}(x)$, for the special case $N_i = N_c /
l$, where
\eqn\ancoverl{{a _{IR}(x) \over N_f^2} = {1 \over 9} \left (4 + {16
\over l}\right )x^2 + {4 \over 9}x.}
Comparing \amaxi\ and
\ancoverl, we see that $a_{\widehat O}(x)>a_{IR}(x)$ is indeed
satisfied for
all $x>1$. Figure 5 shows the result for the case $l=2$.

\bigskip
\centerline{\epsfxsize=0.50\hsize\epsfbox{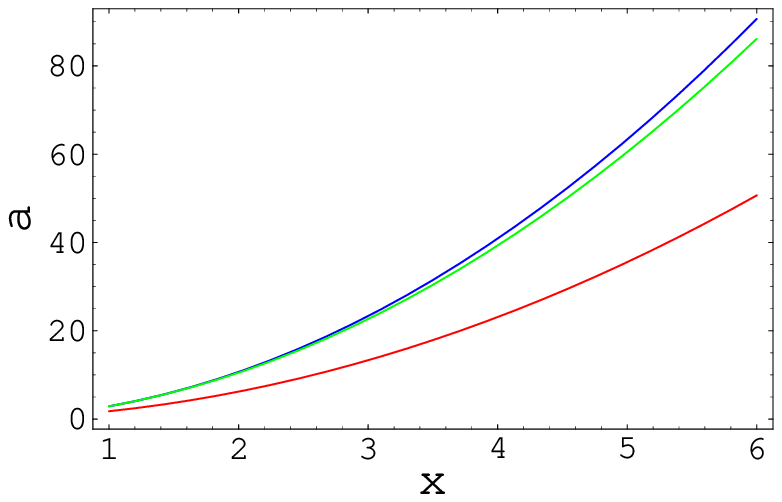}}
\centerline{\ninepoint\sl \baselineskip=8pt {\bf Figure 5:}
{\sl $N_f^{-2}a_{free}(x)$ (top, blue), $N_f^{-2}a_{\widehat O}(x)$
(middle, green),
and $l=2$ broken $N_f^{-2}a_{IR}(x)$ (bottom, red) theories.}}
\bigskip

\subsec{Relevant superpotential deformations of the $\widehat O$
fixed points}

We now classify the relevant superpotential deformations of the $
\widehat O$ RG fixed points, of the form $W=\Tr X^nY^p$.
This is  relevant if $\Delta (W)<3$, i.e. if $R(W)=(n+p)z(x)<2$.
Because $z(x)$ is a monotonically decreasing function of $x$,
asymptoting to $z(x\rightarrow \infty) \rightarrow \half$, the only
relevant possibilities for all $x\geq 1$ are quadratic and cubic
superpotentials.
The only such independent relevant deformations are
\eqn\oreld{\eqalign{W&=\Tr XY:\ \widehat O\rightarrow SQCD,\cr
W&=\Tr Y^2:\ \widehat O\rightarrow \widehat A,\cr
W&=\Tr XY^2:\ \widehat O\rightarrow \widehat D, \cr
W&=\Tr Y^3:\ \widehat O\rightarrow \widehat E.}}
All of these are relevant for all $x$ in the range $1<x\leq \infty$.
The deformation $W=\Tr XY$ gives a mass to both adjoints, taking the
theory to SQCD with no adjoints (which does not lead to any RG fixed
point
in the $x>1$ range).  The deformation $W=\Tr Y^2$ gives a mass to one
of the adjoints, driving the theory to the $\widehat A$ RG fixed points
of SQCD with one adjoint $X$ and $W(X)=0$; this
$\widehat A$ RG fixed point was analyzed in detail in \KPS. The
remaining
relevant deformations in \oreld\ drive the $\widehat O$ RG fixed points
to new RG fixed points, which we name $\widehat D$ and $\widehat E$
and discuss further in the following sections.

\newsec{The $\widehat E$ RG fixed points: $W _{\widehat E}=\Tr Y^3$}

We have seen that $W_{\widehat E}=\Tr Y^3$ is a relevant deformation
of the $\widehat O$ SCFTs for all $x>1$, driving  $\widehat O$
to some new RG fixed points which we name
$\widehat E$.  The $\widehat E$ SCFTs exist for all $x>1$.
When the theory is just barely asymptotically
free, $x=1+\epsilon$ with $0<\epsilon \ll 1$, the $\widehat
O\rightarrow \widehat
E$ RG flow can be analyzed
perturbatively; this is discussed in Appendix A. Upon increasing $x$ the
coupling of these SCFTs becomes stronger. We
use a-maximization to find the exact superconformal
$U(1)_R$ symmetry for all $x\geq 1$.

\subsec{The chiral ring of the $\widehat{E}$ theory.}

The gauge invariant chiral operators of the $\widehat{E}$ theory are a
subset of those of the $\widehat{O}$ theory, where we impose chiral
ring relations coming from the superpotential  $W=\Tr Y^3$.  For
$U(N)$, this yields $Y^2=0$ in the chiral ring, while for $SU(N)$ it
gives
$Y^2=\alpha {\bf 1}$; $\alpha$ is a Lagrange multiplier used
to set $\Tr Y=0$.  For convenience, we'll consider the simpler $U(N)$
ring relation; for large $N$ there isn't much difference in any case.

Imposing $Y^2=0$ in the ring, we can form operators such as
\eqn\ehatcr{\O_{I_1 \dots I_n} = \Tr X_{I_1} \cdots X_{I_n},}
where, $I_i$=1,2 labels the two adjoints, as long as no two adjacent
(including via trace cyclicity) adjoints are $Y$'s.  Such operators can
still have many net $Y$'s, e.g. $\Tr (XY)^{53}$.
Similarly, the meson and baryon
operators are the subset of those in sect. 2. for which there are no
two adjacent $Y$'s, e.g.
$\widetilde Q_{\tilde i} Y(XY)^{29}Q_i$.

\subsec{Finding the superconformal $U(1)_R$ via a-maximization}

At the eventual IR fixed point controlled by the superpotential $W=\Tr
Y^{3}$, we impose $z={2\over 3}$ in \rtwoadj, to ensure $R(W)=2$,
yielding the 1-parameter family
\eqn\rwithw{R(Y)={2 \over 3}, \qquad R(Q)=R(\widetilde{Q})\equiv
y \qquad
R(X) = {1+x-y \over x} - {2 \over 3}.}
The superconformal $U(1)_R$ is determined by maximizing the
central charge $a$, given by \atwoadj\ with
$z=2/3$, with respect to $y$; the result is
\eqn\ywone{y(x)=1+{x(2-\sqrt{10x^2-1}) \over
3(2x^2-1)}.}
This result is monotonically decreasing with $x$, with asymptotic value
as
$x\rightarrow \infty$
\eqn\relim{y(x\rightarrow \infty) \rightarrow 1-{\sqrt{10}\over
6}\approx .4730,
\qquad R(X)\rightarrow {1\over 3}.}
In figs. 6 and 7, we plot  $y(x)$ and $R(X)(x)$ for this theory.

\bigskip
\centerline{\epsfxsize=0.50\hsize\epsfbox{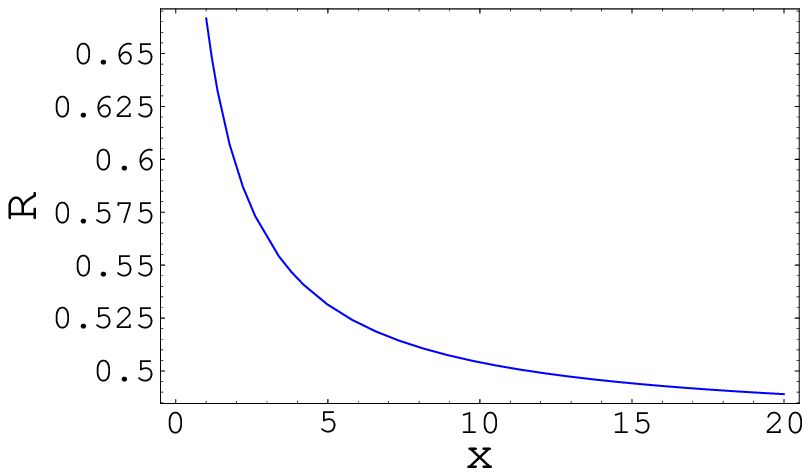}}
\centerline{\ninepoint\sl \baselineskip=8pt {\bf Figure 6:}
{\sl $y(x)$ for the $\widehat{E}$ theory.}}
\bigskip

\bigskip
\centerline{\epsfxsize=0.50\hsize\epsfbox{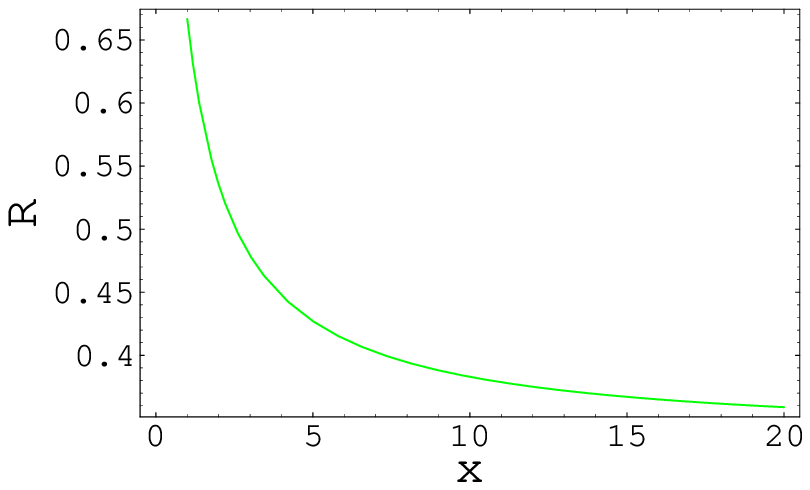}}
\centerline{\ninepoint\sl \baselineskip=8pt {\bf Figure 7:}
{\sl $R(X)(x)$ for the $\widehat{E}$ theory.}}
\bigskip

Since $y>0.47$ and $R(X)>1/3$ for all $x$, no gauge invariant chiral
operator ever
violates the unitarity bound $R\geq 2/3$  (aside from $\Tr X$ and $\Tr
Y$ in the
$U(N)$ version, but they make negligible contribution for large $N$ and
$N_f$.)
So unitarity does not
require any accidental symmetries, and we tentatively
propose that the above results are correct as they stand for all $x>1$.
Again, it is possible that for sufficiently large $x$ the strongly
coupled
theory actually does have some accidental symmetries.  But we do not see
any indication of this possibility in our results.

Plugging $y(x)$ given by \ywone\ and $z=2/3$ into the expression
\atwoadj\ yields
\eqn\aehatis{{a_{\widehat E}(x)\over N_f^2}=
{ 2x^2(17+36x^4-66x^2 +(10x^2-1)^{3 \over 2}) \over 9(1-2x^2)^2 },}
which we have plotted in fig. 8, together with the
$N_f^{-2}a_{\widehat O}(x)$ found in the previous section.  Since
perturbing $\widehat O$ by $W_{\widehat E}$ induces the RG flow
$\widehat O\rightarrow \widehat E$,
the conjectured a-theorem prediction is $a_{\widehat O}(x)>a_{\widehat
E}(x)$ for all $x$; this is indeed seen to be satisfied in fig. 8.  We
could have
anticipated this because $a_{\widehat O}(x)$ involved maximizing
with respect to $z$,
whereas in $a_{\widehat E}(x)$ it was constrained to $z={2\over 3}$
(though one must generally be careful with this argument, because the
maxima are only local ones) \IW.

\bigskip
\centerline{\epsfxsize=0.50\hsize\epsfbox{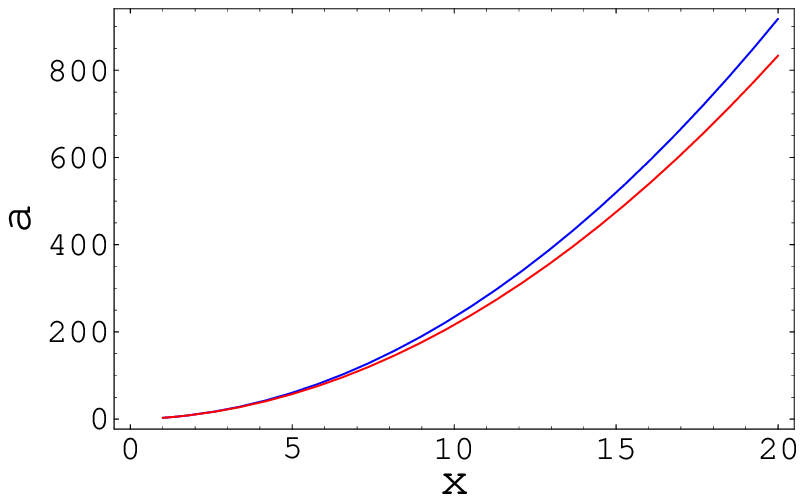}}
\centerline{\ninepoint\sl \baselineskip=8pt {\bf Figure 8:}
{\sl $a/N_f^2$ for the $\widehat{O}$ (top, blue) and $\widehat{E}$
(bottom, red) theories.}}
\bigskip

\subsec{Relevant deformations of the $\widehat E$ RG fixed points}

We consider deforming the superpotential as $W=\Tr Y^3\rightarrow
W=\Tr Y^3+\Delta W$, with $\Delta W$ of
the general form $\Delta W=\Tr X^n
Y^p$ (allowing for various inequivalent $X$ and $Y$ orderings).  This
$\Delta W$ will
be relevant, and can then lead to new SCFTs, if
\eqn\reek{R(\Delta W=X^nY^p)={2p\over 3}+n\left({1\over 3}+{1-y(x)\over
x}\right)<2.}
Since $R(X)$ is monotonically decreasing,
the highest possible $p$ and $n$ are found by considering the
$x\rightarrow \infty$ limit of \reek, where $R(X)\rightarrow {1\over
3}$:
\eqn\rekrel{2p + n <6.}
This leads to the several possibilities, which we now discuss.

First, any quadratic $\Delta W$ superpotential is relevant.  The
possibilities, and where they drive the $\widehat E$ SCFTs, are
\eqn\edefq{\eqalign{\Delta W&=\Tr Y^2:\ \widehat E\rightarrow \widehat
A,
\cr
\Delta W&=\Tr X^2:\ \widehat E\rightarrow A_1, \cr
\Delta W&=\Tr XY:\ \widehat E\rightarrow \rm{SQCD}.}}
These flows are all consistent with the a-theorem; e.g.
in fig. 9 we see that $a_{\widehat E}(x)>a_{\widehat A}(x)$ is
satisfied.

\bigskip
\centerline{\epsfxsize=0.50\hsize\epsfbox{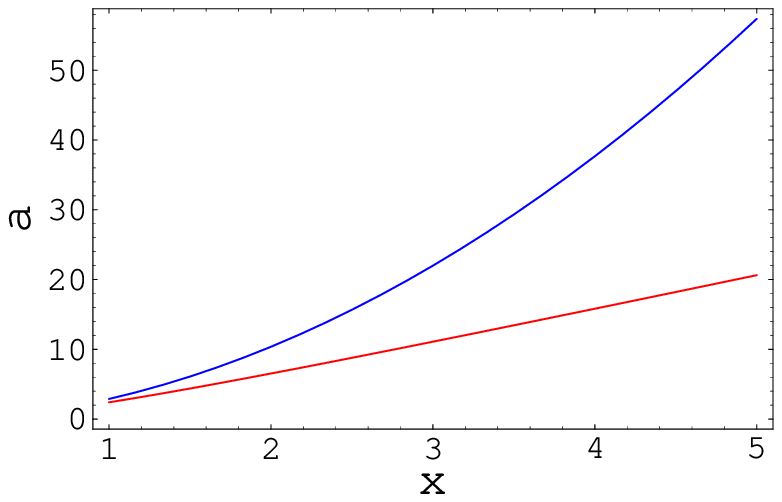}}
\centerline{\ninepoint\sl \baselineskip=8pt {\bf Figure 9:}
{\sl $a/N_f^2$ for the $\widehat{E}$ (top, blue) and $\widehat{A}$
(bottom, red) theories.}}
\bigskip

At the level of cubic $\Delta W$, the only independent, relevant
possibility is
\eqn\edefc{\Delta W=\Tr X^2Y:\ \widehat E\rightarrow D_4,}
in the terminology of \oade.  Deforming $\Tr Y^3$ by $\Delta W=\Tr X^3$
is equivalent to \edefc\ via a change of
variables, and $\Tr XY^2$ is eliminated by the $\widehat E$ chiral ring
relation.
Using \reek\ and the results of the previous subsection, we can see
that $\Tr X^2Y$ is relevant for all $x> 1$ (since
$y(x)>1-{1\over 3}x$ for all $x>1$).
Though $\Delta W=\Tr X^2Y$ is relevant, we do not expect
that it ever wins out over the original $W_{\widehat E}=\Tr Y^3$ term;
both are important in determining the
eventual RG fixed point.  This will be further discussed in
Appendix A.

Finally, we have the higher-degree $\Delta W$ solutions of \reek.
These are only relevant if $x$ is sufficiently large, and the
independent
possibilities (easily seen from \rekrel) for
$W=W_{\widehat E}
+\Delta W$ are:
\eqn\revi{\widehat E\rightarrow E_6:\ W_{E_6}=\Tr (Y^3+X^4)
\qquad \hbox{if}\quad x\geq x^{min}_{E_6}
\approx 2.55,}
\eqn\revii{\widehat E\rightarrow E_7:\ W_{E_7}=\Tr (Y^3+YX^3)\qquad
\hbox{if}\quad x\geq x^{min}_{E_7}\approx 4.12,}
\eqn\reviii{\widehat E\rightarrow E_8:\ W_{E_8}=\Tr (Y^3+X^5)\qquad
\hbox{if}\quad x\geq x^{min}_{E_8}
\approx 7.28.}
The values of $x^{min}_{E_6}$, $x^{min}_{E_7}$, and $x^{min}_{E_8}$
are obtained by plotting, as in fig. 10, the R-charge \reek\ of the
corresponding
deformation, $\Delta W_{E_6}=\Tr X^4$, $\Delta W_{E_7}=\Tr YX^3$, and
$\Delta W_{E_8}=\Tr X^5$, at the $\widehat E$ RG fixed point,
and seeing when $R(\Delta W)$ just drops below $R=2$, i.e.
when the inequality in \reek\ is saturated.
\bigskip
\centerline{\epsfxsize=0.50\hsize\epsfbox{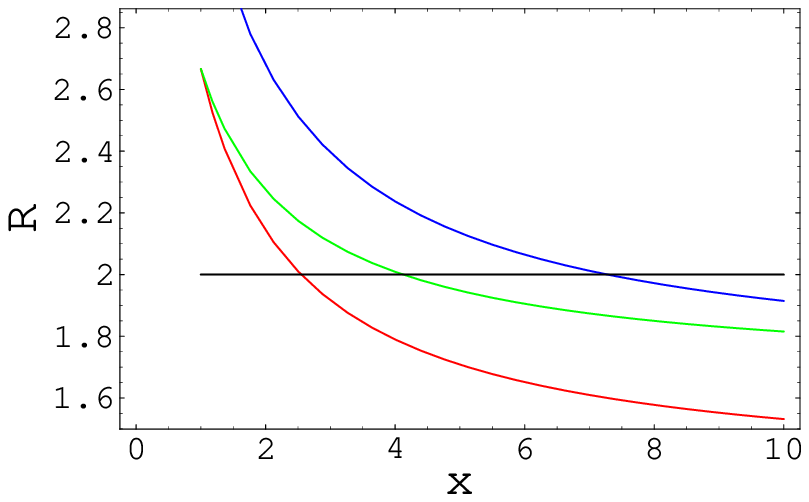}}
\centerline{\ninepoint\sl \baselineskip=8pt {\bf Fig. 10:}
{\sl $R(\Delta W_{E_8})$ (top,blue), $R(\Delta W_{E_7})$ (middle,
green),
and $R(\Delta W_{E_6})$ (bottom, red), in the $\widehat{E}$ theory.}}
\bigskip
If $x>x_{E_{6,7,8}}^{min}$, these $\Delta W$ drive $\widehat E$ to new
SCFTs,  which we call $E_6$, $E_7$, and $E_8$ and will analyze
further in sect. 7.

\newsec{The $\widehat D$ RG fixed points, $W_{\widehat D}=\Tr XY^2$}

We have seen that $W_{\widehat D}=\Tr XY^2$ is a relevant deformation
of the
$\widehat O$ SCFTs, driving them to new SCFTs, which we name $\widehat
D$, for all $x\geq 1$.
When the theory is just barely asymptotically free, $x=1+\epsilon$ with
$0<\epsilon \ll 1$, the flow from the $\widehat O$ RG fixed points to
the $\widehat D$ RG fixed points can be studied in perturbation theory,
as will be done in Appendix A.  For larger $x$ the $\widehat D$ SCFT
becomes more and more strongly coupled.  In this section we
exactly determine the superconformal R-charges and central charge $a$
for all $x$.

\subsec{The chiral ring for the $\widehat{D}$ theory}

The chiral ring of gauge invariant operators is now subject to the
relations
coming from the $W_{\widehat D}$ EOM.  For $U(N_c)$
these are $\partial _YW_{\widehat D}=\{ X, Y \} =0$ and $
\partial _XW_{\widehat D}=Y^2=0$; again, for $SU(N_c)$
there would be Lagrange multipler unit matrices on the RHS, which
we ignore in any case.  The relation $\{X,Y\}=0$ in the chiral ring
is particularly convenient, since it allows us to freely re-order $X$
and $Y$
superfields (up to a minus sign). Using these ring relations, the only
non-zero operators of the form \oops\ are
\eqn\dhatcr{\Tr X^l, \qquad l\geq 0,}
(for $U(N)$ we also should include $\Tr Y$).   Note that $\Tr X^nY=0$
using $\{ X,Y\} =0$ and cyclicity of the trace.  The non-zero mesons are
\eqn\dhatmes{M_{\ell, j}=\widetilde{Q}_{\widetilde{i}}X^\ell Y^jQ_i
\quad
\hbox{for}\quad \ell \geq 0, \quad j=0,1.}
The non-zero baryons are
\eqn\dhatbary{Q^{n_{(0,0)}}_{(0,0)}Q^{n_{(1,0)}}_{(1,0)}\cdots
Q^{n_{(l,0)}}_{(l,0)}Q^{n_{(0,1)}}_{(0,1)}Q^{n_{(1,1)}}_{(1,1)}\cdots
Q^{n_{(k,1)}}_{(k,1)}}
where
\eqn\dhatdq{Q_{(l,j)} = X^l Y^j Q, \qquad l\geq 0, \quad j=0,1,}
and
\eqn\dbaryrest{\sum_{j=0}^l n_{(j,0)} + \sum_{j=0}^k n_{(k,1)} = N_c,
\quad
n_{(l,j)} \leq N_f, \quad l,k \geq 0}

\subsec{$a$-maximization, this time with accidental symmetries}
At the $\widehat D$ RG fixed point, by $W_{\widehat D}=\Tr XY^2$,
there is a one-parameter family of anomaly free R-charges
satisfying $R(W_{\widehat D})=2$:
\eqn\rxysquared{R(Q)=R(\widetilde{Q}) = y, \quad R(Y)= {y-1 \over x} +
1,  \quad
R(X)={2-2y \over x}.}  Plugging these into \atrial,
we see that $y$ is determined by maximizing
\eqn\axysquared{\eqalign{{a _{\widehat D}^{(0)}(x)\over N_f^2} &=
2x^2+3x^2\left({y-1 \over x}\right )^3
+3x^2\left ( {2 -2y \over x} -1 \right )^3 \cr & \quad - x^2\left ( {y-1
\over x} \right )
-x^2\left ( {2 -2y \over x} -1 \right ) + 6x(y-1)^3 -2x(y-1).}}
This is maximized for
\eqn\yxysquared{y^{(0)}(x)
= 1+{x(12-\sqrt{11+38x^2})\over 3(2x^2-7)}.}
The superscript on $a^{(0)}$ and $y^{(0)}$ will be explained presently.

Unlike the previous cases, now we do see that $y^{(0)}(x)$, as given by
\yxysquared, would lead to a unitarity bound violation if it were
extrapolated to large $x$.  Indeed,
$y^{(0)}(x)$ \yxysquared\ approaches a negative number at large $x$:
$y^{(0)}(x\rightarrow \infty)\rightarrow 1-{1\over 6}\sqrt{38}\approx
-0.027$.
The expressions \axysquared\ and \yxysquared\ are correct only in the
range of $x$ given by $1\leq x\leq x_1$, where $x_1$ is
where $R(\widetilde QQ)=2y={2\over 3}$, where
the first meson crosses the unitarity bound.  $x_1$ is thus
the solution of $y^{(0)}(x_1)=1/3$, giving
$x_1\approx 3.67$.  For $x>x_1$, we need to re-work the above
procedure, taking into account the free meson $M$ and corresponding
accidental symmetry; how to do this was found in \KPS\ (in the context
of the $\widehat A$ SCFTs) and is reviewed in appendix B.

As we continue to increase $x$, more and more generalized mesons,
$(M_{p,0})_i^{\widetilde i}=\widetilde Q^{\widetilde i} X^{p-1}Q _i$,
hit
the unitarity bound and then become free fields.  Mesons involving the
operator $Y$, $M_{p,1}\equiv \widetilde Q X^{p-1} YQ$ never violate the
unitarity bound since, as seen from
\rxysquared, $R(Y)$ is always rather large: $R(Y)\geq 1$ and $R(X)\geq
0$, with $R(Y)\rightarrow 1$ for $x\rightarrow \infty$.  As will be
shown in appendix C, for all $x$, no baryons ever hit the unitarity
bound. The chiral ring elements \dhatcr\ will hit the unitarity bound
and
become free, but we can ignore their contribution in the large $N_f$
limit,
since they are down by a factor of $N_f^2$ as compared with the meson
contributions.    So, for all $x$, we only need to account for the
mesons
$(M_p)_i^{\widetilde i}=\widetilde Q^{\widetilde i} X^{p-1}Q _i$ hitting
the unitarity bound and becoming free; this happens successively in $p$
as we increase $x$.

Let $x_p$ be the value of $x$ where the $N_f^2$ mesons
$M_p=\widetilde
Q X^{p-1}Q$ hit the unitarity bound:
\eqn\pmesx{2y(x_p)+(p-1){2-2y(x_p)\over x_p}={2\over 3}.}
For $x$ in the range $x_p\leq x \leq x_{p+1}$, the mesons $M_{\ell,0}$
with $\ell =1\dots p$ are free fields, while those with $\ell >p$ are
interacting.
We account for the accidental symmetries of the free mesons $M_1\dots
M_p$
by using the modified central charge $a^{(p)}$,
which is given as in \KPS\ by
\eqn\apis{\eqalign{{a _{\widehat D}^{(p)}(x)\over N_f^2} &=
2x^2+3x^2\left({y-1 \over x}\right )^3
+3x^2\left ( {2 -2y \over x} -1 \right )^3 \cr & \quad - x^2\left ( {y-1
\over x} \right )
-x^2\left ( {2 -2y \over x} -1 \right ) + 6x(y-1)^3 -2x(y-1)\cr
&+{1\over 9}
\sum _{j=0}^{p-1}
\left[2-3(2y+j{2-2y\over x})\right] ^2
\left[5-3(2y+j{2-2y\over x})\right].}}
Maximizing the function \apis\ with respect to $y$ yields the function
$y^{(p)}(x)$.

The R-charge $R(Q)\equiv y(x)$ is given by patching together
these various functions:
\eqn\ydpatch{y(x)=y^{(p)}(x) \qquad \hbox{for}\qquad x_p<x<x_{p+1}}
and the central charge is given by patching together the maximal values
of the \apis:
\eqn\adpatch{{a_{\widehat D}(x)\over N_f^2}={a_{\widehat D}
^{(p)}(x)\over N_f^2}
\qquad \hbox{for}\qquad x_p<x<x_{p+1}.}
When we solve for $x_p$, we use \pmesx\ with $y(x_p)=y^{(p-1)}(x_p)$,
and
iterate this procedure to all $x$.
The functions $y(x)$ and $a_{\widehat D}(x)$ defined by \ydpatch\ and
\adpatch\ are continuous and smooth, as in \KPS, despite the patching.
We plot
the resulting $y(x)$ and $N_f^{-2}a_{\widehat D}(x)$
(obtained numerically) in figs. 11 and 12. In fig. 12, we plot
$N_f^{-2}a_{\widehat D}(x)$ along with $N_f^{-2}a_{\widehat O}(x)$ so
that one may verify that the a-theorem is satisfied:
$a_{\widehat{O}}(x) > a_{\widehat{D}}(x)$ for all $x>1$.
\bigskip
\centerline{\epsfxsize=0.50\hsize\epsfbox{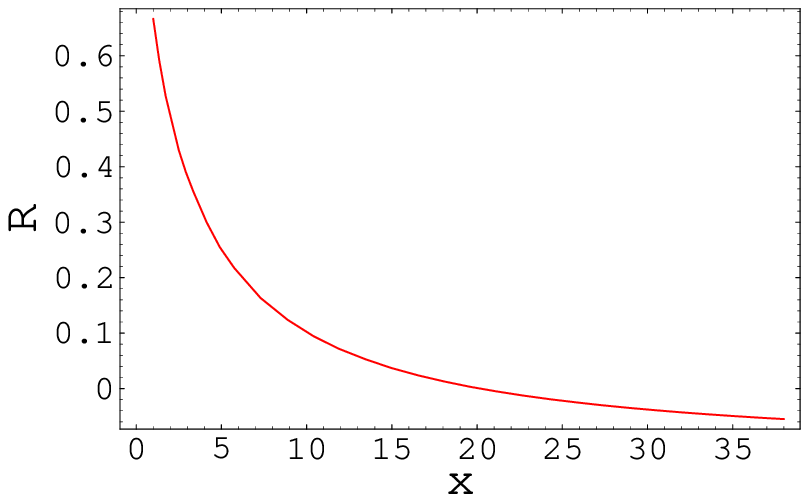}}
\centerline{\ninepoint\sl \baselineskip=8pt {\bf Figure 11:}
{\sl $y(x)$ for the $\widehat{D}$ theory.}}
\bigskip

\bigskip
\centerline{\epsfxsize=0.50\hsize\epsfbox{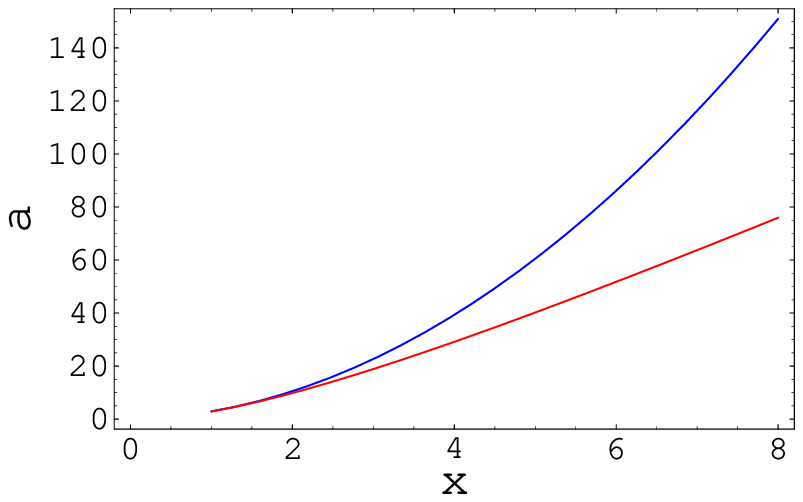}}
\centerline{\ninepoint\sl \baselineskip=8pt {\bf Figure 12:}
{\sl $a_{\widehat{O}}/N_f^2$ (top, blue) and
$a_{\widehat{D}}/N_f^2$ (bottom, red).}}
\bigskip

We can analytically solve for the asymptotic $x\gg 1 $ behavior of the
$y(x)$ obtained by this procedure (in analogy with the $\widehat A$
case in \KPS).  In the large $x$ limit, we can replace the sum over
$j$ with an integral over $v$, defined by
\eqn\vis{v=2-3\left(2y+(j-1){2-2y\over x}\right).}
In the $x\gg 1$ limit (where $v$ becomes a continuous variable), this
yields
\eqn\aplarge{{a_{\widehat D}(x,y)\over N_f^2}\approx {a_{\widehat
D}^{(0)}(x,y)
\over N_f^2}+
\int _0^{2-6y}dv{xv^2(3+v)\over 54(1-y)}
\approx x\left(6(y-1)^3-19(y-1)+{2\over
9}(
1-3y)^3\right),}
which is maximized for
\eqn\yasymp{y(x\rightarrow \infty) = -1/8.}
The asymptotic value for the
central charge is then
$${ a_{\widehat D}(x)\over N_f^2}\approx {1931\over 144}x\approx 13.41x
\qquad\hbox{for large $x$}.$$

\subsec{Relevant deformations}

In the limit $x\rightarrow \infty$, we see from \rxysquared\ and
\yasymp\
that $R(X)\rightarrow 0$.  Thus a deforming superpotential $\Delta W
=\Tr X^{k+1}$ of the $\widehat D$ RG fixed point will be relevant for
any
$k$, provided that $x$ is chosen sufficiently large.  This is because
\eqn\rdki{R(X^{k+1})=2(k+1){(1-y(x))\over x}\leq 2}  will always be
satisfied
for $x$ larger than some critical value $x^{min}_{D_{k+2}}$ where the
inequality \rdki\ is saturated.    Thus
\eqn\ddkflow{\Delta W=\Tr X^{k+1}: \widehat D\rightarrow D_{k+2} \qquad
\hbox{if}\quad x>x^{min}_{D_{k+2}}.}

The superpotential $\Tr X^{k+1}$ for the case $k=2$
is a relevant deformation of the $\widehat D$ SCFTs, driving them
to the $D_4$ SCFT for all $x\geq 1$.
In particular, this flow can be analyzed in the perturbative regime
$x=1+
\epsilon$, with $0<\epsilon \ll 1$, as is discussed in Appendix A.
Increasing $k$ leads to larger and larger values of $x^{min}_{D_{k+2}},$
so we need to be careful to use the appropriate $y^{(p)}(x)$,
determined via
\apis , in \rdki.  E.g. for $k$ sufficiently small so that
$x^{min}_{D_k}$ is below the value $x_1 \approx 3.67$
where the meson $\widetilde{Q}Q$ becomes free,
we can use \yxysquared. This gives
\eqn\xdkminex{x_{D_{k+2}}^{min}={\sqrt{10 - 34k + 19k^2} \over
3\sqrt{2}}
    \quad\hbox{for}\quad k< 5.}
The first few are
\eqn\xdkminfew{x_{D_5}^{min}=2.09, \quad x_{D_6}^{min}=3.14,
\quad x_{D_7}^{min}=4.24, \quad x_{D_8}^{min}=5.37,}
for $k=3,4,5,6$.

For large $k$, the $x_{D_{k+2}}$ become
large and we can use the asymptotic value
$y(x\rightarrow \infty )=-1/8$
to get \eqn\xdklg{x^{min}_{D_{k+2}}\rightarrow {9\over 8}k
\qquad \hbox{for}\qquad k\gg 1.}  Also (in analogy with the $A_k$
case discussed in \KPS), we have the general inequality
\eqn\xdkminin{{x_{D_{k+3}}^{min}\over k+2}>{x_{D_{k+2}}^{min}\over
k+1},}
which follows from \rdki, which gives
$x_{D_{k+2}}^{min}/(k+1)=1-y(x_{D_{k+2}}^{min})$,
together with the fact that $y(x)$ is monotonically decreasing in $x$.
Using $y(x\rightarrow \infty)= -1/8$ we get the inequality
\eqn\xdkminii{{x_{D_{k+2}}^{min}\over k+1}<1-y(x\rightarrow
\infty)={9\over
8},}
which is saturated \xdklg\ for $k\rightarrow \infty$.   Even for low
$k$, this
estimate isn't too far off, e.g. it would give for $k=6$:
$x_{D_8}^{min}<7.875$, which isn't so far off from \xdkminfew.

\newsec{The $D_{k+2}$ RG fixed points: $W=\Tr (X^{k+1}+XY^2)$.}

Consider deforming the $\widehat O$ RG fixed points by the
superpotential
\eqn\dkwg{W=\lambda _1\Tr X^{k+1}+\lambda _2 \Tr XY^2.}
If we were to start with $\lambda _2=0$, we have already seen that the
$\lambda _1$ deformation would only be relevant for $k\leq 2$.  But if
we take $\lambda _2\neq 0$, the theory first flows to be near the
$\widehat D$ RG fixed point.  Then, starting at the $\widehat D$ RG
fixed point, we have seen in the previous section that $\Tr X^{k+1}$
is relevant for all $k$, provided that $x>x_{D_{k+2}}^{min}$.  When
$x>x_{D_{k+2}}^{min}$ we thus expect that the RG flow is as in
Fig. 13.  If $x<x_{D_{k+2}}^{min}$ the flow is instead as in Fig. 14.
\bigskip
\centerline{\epsfxsize=0.50\hsize\epsfbox{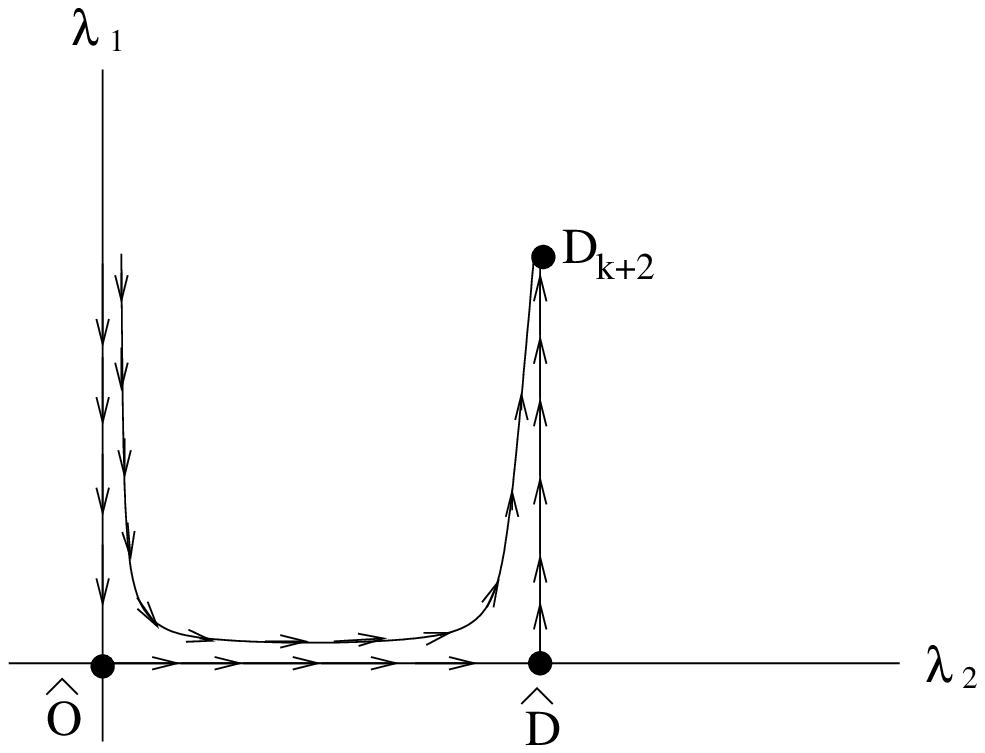}}
\centerline{\ninepoint\sl \baselineskip=8pt {\bf Figure 13:}
{\sl RG flow in the $\lambda_1-\lambda_2$ plane when
$x>x_{D_{k+2}}^{min}$.}}
\bigskip

\bigskip
\centerline{\epsfxsize=0.50\hsize\epsfbox{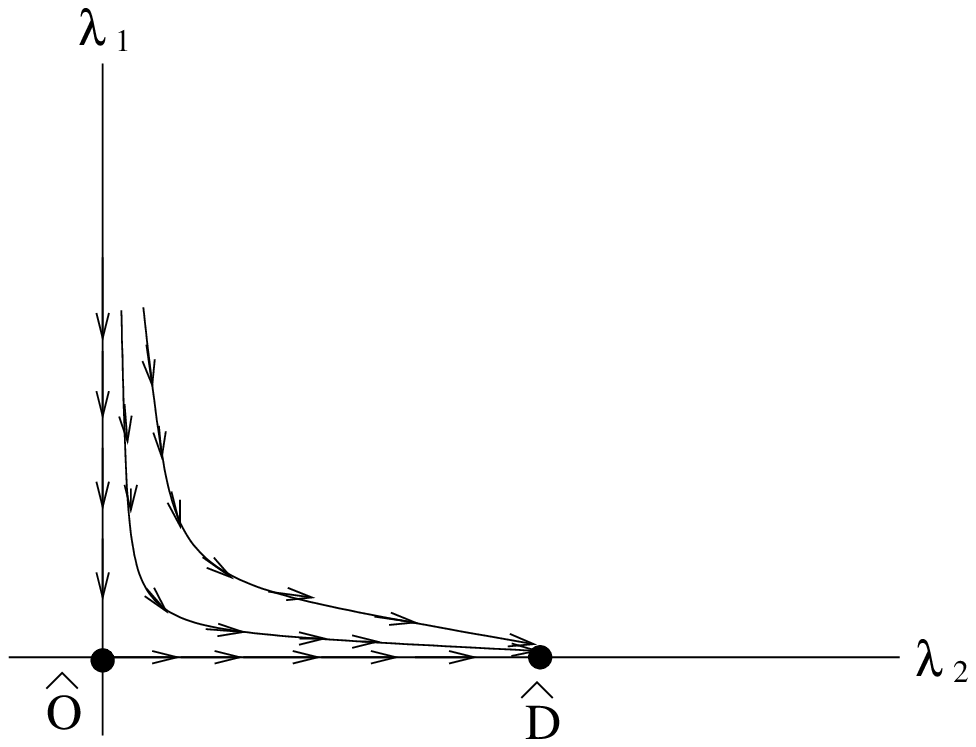}}
\centerline{\ninepoint\sl \baselineskip=8pt {\bf Figure 14:}
{\sl RG flow in the $\lambda_1-\lambda_2$ plane when
$x<x_{D_{k+2}}^{min}$.}}
\bigskip

At the $D_{k+2}$ SCFT fixed point, $\lambda _1\rightarrow \lambda
_{1*}$ and
$\lambda _2\rightarrow \lambda _{2*}$.  We will rescale $X$ and $Y$ to
absorb these coefficients into the Kahler potential, taking
$W_{D_{k+2}}$ as in
\oade.

\subsec{The chiral ring, and the stability bound}

As before, the classical chiral ring is that of the $\widehat O$ theory,
subject to the additional relations coming from the EOM of
$W_{D_{k+2}}$: $\{X,Y\}=0$ and $X^k +Y^2=0$
in the chiral ring.  The result for $k$ odd is very different from the
$k$ even case.

For $k$ odd, these imply that $Y^3=0$ in the chiral
ring.  To see this (see also \refs{\JB, \CKV}),
multiply the second equation of motion
by $Y$  on the left and add this to the same equation
multiplied by $Y$ on the right, to get $YX^k + X^kY= -2Y^3$.
Anticommuting all $Y$ fields to the right then gives $((-1)^k +
1)X^kY=-2Y^3$ and
hence $Y^3=0$ for $k$ odd.  The independent non-zero products of $X$ and
$Y$ are then truncated to
\eqn\dktrunc{X^{\ell -1}Y^{j-1}, \qquad \ell =1\dots k, \quad j=1\dots
3,}
where the order of $X$ and $Y$ does not matter because of the chiral
ring relation $\{X,Y\}=0$.
When we take the traces to form gauge invariant operators, the only
non-zero ones (due to $\{ X,Y\} =0$ and cyclicity of the trace) are the
$k+2+\half (k-1)$ operators
\eqn\dkoops{\Tr X^{\ell -1} \quad (\hbox{for}\ \ell =1\dots k),\qquad
\Tr Y, \qquad
\Tr Y^2, \qquad \Tr X^{2n}Y^2 \quad(\hbox{for}\ n=1\dots \half (k-1)).}
We can also form the $3kN_f^2$ mesons
\eqn\dkmesons{M_{lj} = \widetilde{Q} X^{l-1}Y^{j-1} Q; \quad l=1,\cdots
, k ; \quad j=1,2,3,}
and the baryons
\eqn\dkbary{B^{(n_{1,1},n_{2,1},\cdots,n_{k,3})}=Q^{n_{1,1}}_{(1,1)}
\cdots
Q^{n_{k,3}}_{(k,3)}.
\quad \sum_{l=1}^k \sum_{j=1}^3 n_{l,j} = N_c; \quad n_{l,j}\leq N_f}
formed from the dressed quarks
\eqn\dkquark{Q_{(l,j)} = X^{l-1}Y^{j-1}Q; \quad l=1,\cdots , k ; \quad
j=1,2,3.}

We saw already that the $D_{k+2}$ SCFTs can only exist if
$x>x_{D_{k+2}}^{min}$.
It turns out that these SCFTs, at least for $k$ odd, also have an upper
bound on
the allowed value of $x$:
\eqn\xmaxdk{x<x_{D_{k+2}}^{max}=3k.}
(For  $A_k$ theories the analogous
stability condition is $x<x_{A_k}^{max}=k$ \refs{\DKi,
\DKAS}.)  For $x>x_{D_{k+2}}^{max}$,
the theory is rendered unstable by developing a dynamically generated
superpotential, which will
spoil conformal invariance and drive the theory away from the origin of
the moduli space of vacua.  The previously discussed RG fixed point
theories, $\widehat O$,
$\widehat D$, and $\widehat E$, as well as $\widehat A$, are stable
for all $x$, so for those SCFTs there is no upper bound on $x$.

The stability bound is related to the truncation
to the $3k$ independent products in \dktrunc.  To see the stability 
bound,
deform $W_{D_{k+2}}$ by lower order terms, e.g. to
\eqn\wdkdef{W=\Tr (F_{k+1}(X)+XY^2+\alpha Y),}
where $F_{k+1}(X)$ is a degree $k+1$ polynomial in $X$.
The classical chiral ring relations are the EOM
\eqn\dkceom{XY+YX=-\alpha \quad {\rm and} \qquad Y^2+F'_{k+1}(X)=0.}
The irreducible representations of this algebra were actually
discussed recently in \CKV\ in the context of string theory
realizations of related SUSY gauge theories.  The first relation in
\dkceom\ implies that $X^2$ and $Y^2$ are Casimirs, $[X^2,Y]=0$ and
$[Y^2,X]=0$, so we write $X^2=x^2 {\bf 1}$ and $Y^2=y^2{\bf 1}$.  It
is then seen \CKV\ that the second equation in \dkceom, for $k$ odd,
admits $k+2$ different one-dimensional representations, with $X=x$ and
$Y=y$ for $k+2$ different values of $x$ and $y$.  There are also
$\half(k-1)$ two dimensional representations of the form
$X=\pmatrix{0&a\cr b&0}$ and $Y=\pmatrix{0&c\cr d&0}$.  This is seen
by writing $F'_{k+1}(X)=XP(X^2)+Q(X^2)$,  with the 2d reps specified by
the $\half (k-1)$ roots of $P(ab)=0$.

The $\ev{X}$ and $\ev{Y}$ vacua
solutions of \dkceom\ can have $N_i$ copies of the $i$-th
one-dimensional representation, for $i=1\dots k+2$, and $M_n$ copies
of the $n$-th two-dimensional representation for $n=1\dots \half(k-1)$.
   In
such a vacuum the gauge group is Higgsed as
\eqn\dkhiggs{U(N_c)\rightarrow \prod _{i=1}^{k+2}U(N_i)\prod
_{n=1}^{\half(k-1)}U(M_n)
\qquad\hbox{with}\qquad \sum _{i=1}^{k+2}N_i+\sum _{n=1}^{k-1}2M_n=N_c.}
Each $U(N_i)$ factor in \dkhiggs\ has $N_f$ flavors, while each $U(M_n)$
factor has $2N_f$ flavors.  (It's $2N_f$ because
$U(M_n)=U(M_n)_{diag}\subset U(M_n)\times
U(M_n)\subset U(N_c)$; the $U(M_n)\times U(M_n)$ part is related to an
example considered in
\lref\IntriligatorAX{
K.~A.~Intriligator, R.~G.~Leigh and M.~J.~Strassler,
``New examples of duality in chiral and nonchiral supersymmetric gauge
theories,''
Nucl.\ Phys.\ B {\bf 456}, 567 (1995)
[arXiv:hep-th/9506148].
}
\refs{\IntriligatorAX, \JB}.).  All of the adjoint matter fields get a
mass
for the generic deforming superpotential \wdkdef.
Since each of these low-energy theories is SQCD, the stability
condition is that there is at least one choice of the $N_i$ and $M_n$
such that they all satisfy
$N_i<N_f$ and $M_n<2N_f$.  For this to be the case requires $N_c=
\sum _{i =1}^{k+2}N_i+\sum _{j=1}^{\half(k-1)}2M_j<(k+2)N_f+\half
(k-1)2(2N_f)=3kN_f$.
This is the stability bound \xmaxdk.

The above chiral ring truncation and stability bound were for  $k$ odd.
Classically, there is no such truncation or bound for $k$ even.
Though the classical vacua  and
chiral ring analysis differ for $k$ even vs odd, we
expect that $k$ even and $k$ odd are actually qualitatively similar at
the
quantum level.  For example, we expect that the $D_{k+2}$ SCFTs for
even $k$
have a stability bound similar to \xmaxdk.  The reason is that we can
RG flow from
$D_{k+2}$ fixed points in the UV to $D_{k'+2}$ fixed points in the IR,
for $k'<k$, by deforming the $W_{D_{k+2}}$ superpotential by
$\Delta W=\Tr X^{k'+1}$; in particular we can flow from $k$ odd to $k'$
even. On general grounds, we expect that RG flows always reduce the
stability
bound:
\eqn\xmaxred{x^{max}_{IR}<x^{max}_{UV},}
because the added tree-level
   superpotential terms of the IR theory can only make
it easier to form a dynamically generated superpotential which could
destabilize the origin of the moduli space of vacua.  So we must have
$x^{max}_{D_{k'+2}}<x^{max}_{D_{k+2}}$ for any $k'<k$.  In particular,
if we take $k$ odd and $k'$ even, we see that the $D_{k'+2}$ theory does
have a stability bound.  The simplest possibility compatible
with \xmaxred\ is if \xmaxdk\ applies for all $k$.
For what follows we will mostly specialize to the case of $k$ odd but,
for the reason described above, we expect the $k$ even case to be
qualitatively similar at the quantum level.

\subsec{Computing the central charge $a_{D_{k+2}}(x)$}

We now compute the central charge $a$ for the $D_{k+2}$ theories.
There is no need to employ a-maximization to determine the
superconformal R-charges; they are entirely fixed by the
superpotential $W_{D_{k+2}}$ to be
\eqn\rdkis{R(X)={2\over k+1}; \quad R(Y) = {k \over k+1}; \quad R(Q) =
R(\widetilde{Q})=1-{x \over k+1}.}
As in the $A_k$ case we do, however, still need to account for
the effect of apparent unitarity violations and the associated free
fields in computing the central charge $a$.  Since \rdkis\ implies that
$R(Q)<0$ for $x>k+1$, we will clearly need
to take into account accidental symmetries for the mesons. It naively
appears that baryons could also violate
the unitarity bound for sufficently large $x$; in appendix C we
show that,
much as in the $\widehat A$ case \KPS, no baryons ever actually violate
this bound.

Computing $a$ is straightforward but tedious. Since all mesons
will acquire negative R-charges for large enough $x$, we must figure
out where each of the $3kN_f^2$ mesons \dkmesons\ hits
the unitarity bound. Since the
R-charges are fixed, this is easily done: The R charge of the meson
$M_{lj} = \widetilde{Q} X^{l-1}Y^{j-1} Q$ is given by
\eqn\rmljis{R(M_{lj})=2\left ( 1-{x \over k+1} \right )
+ (l-1){2\over k+1} +(j-1){k \over k+1},}
and this equals 2/3 when
\eqn\xljis{x=x_{lj}\equiv{1\over 6}(-2+k+6l+3kj).}
As one can see, the meson with the largest R-charge, $M_{k3}$, becomes
free at
\eqn\xkiii{x_{k3}={8k-1 \over 3}<3k=x_{D_{k+2}}^{max},}
so, as $x$ increases in the range
$x_{D_{k+2}}^{min}<x<x_{D_{k+2}}^{max}$
eventually all of these mesons become free and we must account for their
accidental symmetries.

As an example, consider the $k=3$ case, with superpotential $W=\Tr X^4
+ \Tr XY^2$. From our results in Section 4.2, we know that in the
range $x<x_{D_5}^{min} \approx 2.09$, the $X^4$ term is not relevant,
and we should use our results for the $\widehat{D}$ theory. At the
point $x_{D_5}^{min}\approx 2.09$, the $X^4$ becomes relevant, and we
can use \rdkis\ for the R-charges. We can then compute the central
charge $a$, being careful to account for the accidental symmetries.
We have plotted $a$ for the $\widehat{D}$ and $D_5$ theories in Figure
15. The two curves touch exactly at the point $x_{D_5}^{min}$ as they
must, since the central charge is a continuous function of $x$.

\bigskip
\centerline{\epsfxsize=0.50\hsize\epsfbox{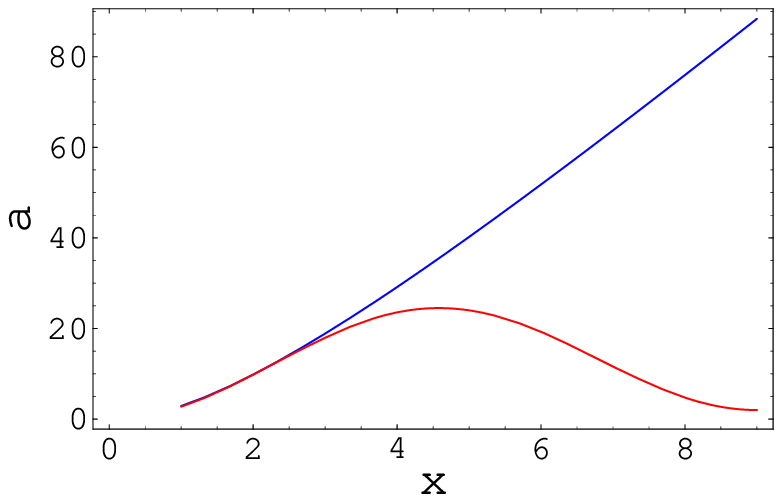}}
\centerline{\ninepoint\sl \baselineskip=8pt {\bf Figure 15:}
{\sl The central charge $a$ for $\widehat{D}$ (top, blue)
and $D_5$ (bottom, red). The curves touch at $x_{D_5}^{min}\approx
2.09$.}}
\bigskip

\subsec{Flow from $D_{k+2}$ to $D_{k^{\prime}+2}$}

We now add a term $X^{k^{\prime}+1}$ with $k^{\prime}<k$ to the
$D_{k+2}$
superpotential.  If $x>x^{min}_{D_{k+2}}$, this is clearly a relevant
deformation,
since we then use the R-charges in \rdkis\ to get
\eqn\rdkdef{R(X^{k^{\prime}+1})=2{k^{\prime}+1 \over k+1}<2.}
If $x<x_{D_{k+2}}^{min}$
the $X^{k+1}$ term is not relevant, so the $X^{k^{\prime}+1}$ is a
deformation of the $\widehat{D}$ theory. As such, it is only relevant
if $x>x_{D_{k^{\prime}+2}}^{min}$, again driving the theory to the
$D_{k'+2}$ SCFT.  For $x<x_{D_{k'+2}}$ both terms are irrelevant
and the theory flows back to the $\widehat D$ SCFT.

As in \refs{\AEFJ,\KPS}, there is a range of $x$ for which
the a-theorem is potentially violated:
\eqn\aviol{a_{D_{k+2}}(x)<a_{D_{k^{\prime}+2}}(x) \quad {\rm for} \quad
1 < x < x_{int}(k+2, k^{\prime}+2).}
For several pairs $(k+2,k^{\prime}+2)$, we have computed this value of
$x_{int}(k+2, k'+2)$:
\eqn\xintdk{\matrix{& x_{int}(8,6)\approx 4.08 \cr
& x_{int}(7,6) \approx 3.64 \cr
& x_{int}(8,5) \approx 3.23 \cr
& x_{int}(7,5) \approx 2.94 \cr
& x_{int}(6,5) \approx 2.56. }}
However, as in the $A_k$ cases \KPS, in no case is the a-theorem ever
actually violated, because all of the above potential violations occur
for
$x$ outside of the range where the $D_{k+2}$ SCFT exists.
Recall from Section 4.2 that $W_{D_{k+2}}$ is only relevant for
$x>x_{D_{k+2}}^{min}$
with
\eqn\xdkminfewii{x_{D_5}^{min}=2.09, \quad x_{D_6}^{min}=3.14,
\quad x_{D_7}^{min}=4.24, \quad x_{D_8}^{min}=5.37.}
Since all of the apparent violations \xintdk\ occur for
$x<x_{int}(k+2,k'+2)$
with $x_{int}(k+2,k'+2)<x_{D_{k+2}}^{min}$ none of them should be
taken seriously: since the $D_{k+2}$ RG fixed point does not exist
for this range of $x$, there is no a-theorem violating RG flow after
all.
In Figure 16, we have plotted what the central charges would have been
if $D_6$ and $D_5$ were relevant for small $x$; one can see that the
a-theorem would potentially be violated for $x<2.56$, but the $D_6$ RG
fixed
point exists only for $x>x_{D_6}^{min}\approx 3.14$.
\bigskip
\centerline{\epsfxsize=0.50\hsize\epsfbox{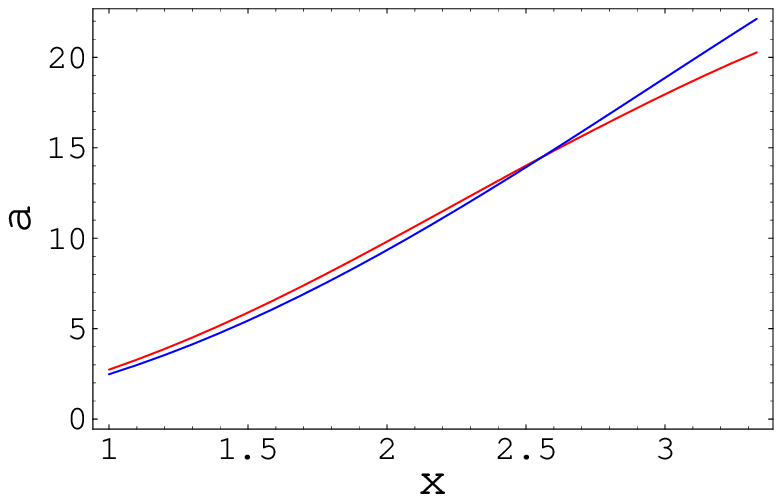}}
\centerline{\ninepoint\sl \baselineskip=8pt {\bf Figure 16:}
{\sl The central charge $a$ for $D_6$ (blue) and $D_5$ (red).}}
\centerline{\ninepoint\sl
The potential violation of the a-theorem is for $1<x<2.56$.}
\bigskip

\newsec{Duality for the $D_{k+2}$ theories.}

A magnetic dual description of the $D_{k+2}$ theories was proposed
in \JB.  We will here discuss and clarify the meaning of this duality.
We also use the duality to determine the behavior of the $D_{k+2}$ RG
fixed
points at large $x$, where there are some accidental symmetries which
are not manifest in the strongly coupled electric description but
are more easily seen in the weakly coupled magnetic dual.

The magnetic dual \JB\ of the $D_{k+2}$ RG fixed point is
an $SU(3kN_f-N_c)$ gauge theory with adjoints $\widetilde X$ and
$\widetilde Y$, $N_f$ magnetic quarks $q_i$ and $\widetilde q^{\tilde
i}$, and $3kN_f^2$ singlets $(M_{\ell j})_{\widetilde i}^i$, $\ell
=1\dots k$,
$j=1,2,3$, with the tree-level superpotential
\eqn\wjbdual{W=\Tr \widetilde X^{k+1}+
\Tr \widetilde X \widetilde Y^2 +
\sum _{\ell =1}^k \sum _{j=1}^3 M_{\ell j}\widetilde q \widetilde
X^{k-\ell }\widetilde Y ^{3-j}q,}
where we omit the RG fixed point coefficients in front of the various
terms in \wjbdual.  We will define $N_{\ell j}\equiv \widetilde q
X^{\ell -1}Y^{j-1}
q$, which are the magnetic mesons.  The superpotential \wjbdual\
implies that
\eqn\mndual{M_{\ell j}\leftrightarrow N_{k+1-\ell, 3-j}}
are Legendre-transform conjugate variables.  Thus, provided that the
corresponding
term in \wjbdual\ is relevant at the RG fixed point, we should include
either $M_{\ell j}$
or $N_{k+1-\ell, 3-j}$ in the spectrum of independent
operators at the RG fixed point, but not both.

The phases of the electric and magnetic theories can be summarized
as
\eqn\phases{\matrix{
x \leq 1 & {\rm free \,\, electric} \cr
1<x<x^{min}_{D_{k+2}} & \widehat{D}\,\, {\rm electric} \cr
x^{min}_{D_{k+2}} < x < 3k-\widetilde{x}^{min}_{D_{k+2}}
& D_{k+2}\,\, {\rm conformal\,\, window} \cr
3k-\widetilde{x}^{min}_{D_{k+2}} < x< 3k-1 & \widehat{D}\,\,
{\rm magnetic} \cr
3k-1 \leq x & {\rm free\,\, magnetic.}}}
\noindent
For $x \leq 1$ the electric theory is not asymptotically free, so it 
flows
to a free theory
in the IR.  In this case, we should definitely use the free-electric
description!  To
see the analogous free-magnetic phase of the magnetic dual,
it's useful to introduce a dual variable to $x$:
\eqn\xmag{\widetilde x \equiv {\widetilde N_c \over N_f}=3k-x.}
The magnetic theory is asymptotically free if $\widetilde x>1$.
When the magnetic theory is not asymptotically free,
i.e. $\widetilde x\leq 1$ and thus $x\geq 3k-1$, the magnetic theory
becomes free
in the IR.  In this case, we should definitely use the magnetic
descrption.  Within the range $1<x<3k-1$, where both electric and
magnetic
theories are asymptotically free, we still have three possibilities.
If $1<x<x_{D_{k+2}}^{min}$
the $\Tr X^{k+1}$ on the electric side is irrelevant, and the electric
theory flows
back to the $\widehat D$ SCFT.  In this case the electric description
is again
definitely better, since it is easier to see the
enhanced symmetries associated with the fact that
$\Tr X^{k+1}$ is irrelevant.  Likewise, in the magnetic theory, the $\Tr
\widetilde X^{k+1}$
superpotential is irrelevant if $\widetilde x<\widetilde
x_{D_{k+2}}^{min}$ (quantities which
we'll discuss shortly) and the magnetic theory then flows to a magnetic
version of the
$\widehat D$ SCFT.  In this case, the magnetic description is
definitely better, since it's
easier there to see the enhanced symmetries associated with the fact
that part of $W_{mag}$
is irrelevant.  Finally, there is a ``conformal window,'' where $\Tr
X^{k+1}$ is relevant
on the electric side, $x>x^{min}_{D_{k+2}}$, and $\Tr \widetilde
X^{k+1}$ is relevant on the magnetic side, $\widetilde x>\widetilde
x^{min}_{D_{k+2}}$.  In
the conformal window, both the electric and the magnetic theories flow
to the
same $D_{k+2}$ SCFT.  Either the electric or the magnetic description
is a useful
description in the conformal window.

The computation of the R-charges for the magnetic theory proceeds
similarly to the analogous computation in \KPS, although here it is
complicated somewhat by the presence of additional fields $M_{\ell j}$.
We consider first the situation where $\widetilde x =1+\widetilde
\epsilon$, with $0<\widetilde \epsilon \ll 1$, so that
the magnetic dual is
just barely asymptotically free.  The magnetic dual theory is then
very weakly coupled and is a very useful description of the IR
physics.  In this small $\widetilde x$ limit only the cubic
terms in \wjbdual, $\Tr \widetilde X\widetilde Y^2+M_{k3}\widetilde
q q$ are relevant; all of the other terms in \wjbdual\ are irrelevant
and can be ignored in the far IR limit.  In particular, for
$k>2$, the $\Tr \widetilde X^{k+1}$ term in \wjbdual\ is irrelevant,
so the magnetic theory actually flows to a $\widehat D$ RG fixed point
rather than a $D_{k+2}$ RG fixed point in the IR.

So for $\widetilde x$ not too far above 1,  the superconformal
$U(1)_R$ charge is given by the magnetic analog of the
$\widehat D$ results \rxysquared:
\eqn\mrxysquared{R(q)=R(\widetilde q)\equiv \widetilde y, \qquad
R(\widetilde Y)={\widetilde y-1\over \widetilde x}+1, \qquad
R(\widetilde X)={2-2\widetilde y\over \widetilde x}.}
The $(3k-1)N_f^2$ fields $M_{lj}$ for all $l\leq k$ and $j\leq 3$,
except for $M_{k3}$ are all decoupled free fields, with $R(M_{lj})=2/3$.
The $N_f^2$ fields $M_{k3}$ couple via the relevant term in the
superpotential
\wjbdual\ and are the Legendre transform conjugate
variable to $N_{11}=\widetilde q q$; the superpotential \wjbdual\ then
fixes
$R(M_{k3})=2-2\widetilde y$.

As we continue to increase $\widetilde x$ from 1, until we reach some
upper bound $\widetilde x_1$,
the only relevant terms in \wjbdual\ are $\Tr \widetilde X^2\widetilde
Y+M_{k3}\widetilde q q$.
We thus compute
\eqn\magaot{\eqalign{{\widetilde a^{(0)}\over N_f^2}&=2\widetilde
x^2+3\widetilde x^2\left({\widetilde y-1 \over \widetilde x}\right )^3
+3\widetilde x^2\left ( {2 -2
\widetilde y \over \widetilde x} -1 \right )^3 \cr & \quad - \widetilde
x^2\left ( {\widetilde y-1
\over \widetilde x} \right )
-\widetilde x^2\left ( {2 -2\widetilde y \over \widetilde x} -1 \right
) + 6\widetilde x(\widetilde y-1)^3 -2\widetilde x(\widetilde y-1)\cr
&+{1\over 9}[(2-3R(N_{11}))^2(5-3R(N_{11})]+{2\over 9}(3k-2),}}
where we define $N_{1,1}\equiv \widetilde q q$ to be the magnetic meson
which is
Legendre transform dual to the interacting meson $M_{3k}$ and
$R(N_{11})=2\widetilde y$.
The first two lines in \magaot\ are simply the magnetic version of
$N_f^{-2}a^{(0)}
_{\widehat D}(\widetilde x, \widetilde y)$, found in \axysquared.  The
last line in
\magaot\ includes the additional contributions of the $(3k-1)N_f^2$ free
field mesons, i.e. all $M_{\ell j}$ aside from $M_{3k}$,
each of which contributes $2/9$ to $a$, along with the contribution of
the interacting meson $M_{k3}$, with R-charge $2-2\widetilde y$.  The
first
term on the last line looks similar to how we would
correct $\widetilde a$ if the meson $N_{11}$ were a free field, but
that is a fake:  we actually should not even include the magnetic meson
$N_{11}$ as an independent field, because the relevant term in
\wjbdual\ makes it
the Legendre transform of $M_{k3}$.  It just happens that the $M_{\ell
j}$
contributions can be written in this similar form to the electric side,
though
the interpretation is different.
We now maximize \magaot\ with respect to $\widetilde y$ to obtain
$\widetilde
y^{(0)}(x)$ and the central charge $N_f^{-2}\widetilde
a^{(0)}(\widetilde x)=
N_f^{-2}\widetilde a^{(0)}(\widetilde x, \widetilde y^{(0)}(\widetilde
x))$.

As  we  continue to increase $\widetilde x$, more of the
previously irrelevant terms in
\eqn\wpiece{\sum _{l=1}^k\sum _{j=1}^3 M_{lj}
\widetilde q \widetilde X^{k-l}\widetilde Y^{3-j}q}
eventually become relevant.  We continue by an iterative procedure.
We compute $N_f^{-2}\widetilde a(\widetilde x,
\widetilde y)^{(p-1)}$
for the theory where $p$ such terms are relevant, with the remaining
$3k-p$
irrelevant.  The $M_{lj}$ entering the relevant terms have R-charge
determined by the superpotential to be
\eqn\dualmesonr{\eqalign{R(M_{lj}) &= 2-2R(q)-(k-l)R(X)-(3-j)R(Y) \cr &=
2-2\widetilde{y}-(k-l)\left ( {2 - 2\widetilde{y} \over \widetilde{x}}
\right ) -
(3-j)\left ( {\widetilde{y} -1 \over \widetilde{x}}+1 \right ), }}
while the $M_{l,j}$ for which the term in \wpiece\ is irrelevant are
free fields,
with $R(M_{lj,})=2/3$.  We then maximize the corresponding
$N_f^{-2}\widetilde a
(\widetilde x, \widetilde y)^{(p-1)}$ with respect to $\widetilde y$ to
    to find $\widetilde y^{(p-1)}(\widetilde x)$ and
$N_f^{-2}\widetilde a^{(p-1)}(\widetilde x)$.

These results for $\widetilde y^{(p-1)}(\widetilde x)$ and $\widetilde
a^{(p-1)}(\widetilde x)$ are applicable until $\widetilde
x>\widetilde x_{p}$, where the next previously irrelevant term
in \wpiece\  becomes relevant, which is when the corresponding
\eqn\rnis{R(N_{k+1-l,3-j})=2\widetilde{y}+(k-l)\left ( {2 -
2\widetilde{y} \over \widetilde{x}}
\right ) + (3-j)\left ( {\widetilde{y} -1 \over \widetilde{x}}+1 \right
)={4\over 3}}
(using $\widetilde y=\widetilde y^{(p-1)}(x)$) for some new values of
$(\ell,j)$.
When this happens, we switch to
\eqn\magacor{{\widetilde a^{(p)}(\widetilde x, \widetilde y)\over
N_f^2}=
{\widetilde a^{(p-1)}(\widetilde x, \widetilde y)\over N_f^2}+{1\over
9}[(2-3R(N_{k+1-l, 3-j}))^2(5-3R(N_{k+1-l,3-j})]-{4\over 9},}
which accounts for the newly interacting field $M_{\ell j}$ having
R-charge
given by
\dualmesonr\ rather than the free-field value $R(M_{kj})=2/3$.   We
then maximize
$N_f^{-2}\widetilde a^{(p)}(\widetilde x, \widetilde y)$ to obtain
$\widetilde y^{(p)}(\widetilde x)$
and $N_f^{-2}\widetilde a^{(p)} (\widetilde x)$ in this next
$\widetilde x$
range, $\widetilde x^{(p)}\leq \widetilde x\leq \widetilde x^{(p+1)}$.
   Note that by the
time any of the magnetic mesons $N_{k+1-\ell, 3-j}=\widetilde q
\widetilde
X^{k-\ell }
\widetilde Y^{3-j}q$ hit their unitarity bound $R(N_{k+1-\ell ,
3-j})=2/3$,
the term involving them in \wpiece\ is already relevant, so
$N_{k+1-\ell ,3-j}$ should not be included as an independent operator,
it
is to be already be eliminated in favor of its Legendre transform field
$M_{\ell j}$.

As we continue to increase $\widetilde x$, eventually we hit
$\widetilde x=\widetilde x_{D_{k+2}}^{min}$, where the term $\Tr
\widetilde X^{k+1}$  becomes relevant:
\eqn\magdxmin{(k+1){2-2\widetilde y(x_{D_{k+2}}^{min})\over
\widetilde x_{D_{k+2}}^{min}}=2.}
The R-charges are then determined without any need for a-maximization,
as in \rdkis. One must still watch out for when mesons
become interacting, since their R-charges are now linearly increasing
functions of $x$. In fig. 17, we have plotted the central charges for
the magnetic $D_5$ theory, for both relevant $\widetilde X^4$ and
irrelevant $\widetilde X^4$.  They touch at the point $\widetilde
x^{min}_{D_5}
\approx 1.86$; for $\widetilde x<1.86$ the term $\Tr \widetilde X^4$ is
irrelevant, while for $\widetilde x>1.86$ it is relevant.  Recall that,
in the electric $D_5$ theory, we found \xdkminfew\ that $\Tr X^4$ is
relevant for $x>x^{min}_{D_{k+2}}\approx 2.09$, so we see that the
critical value for the superpotential to become relevant differs between
the electric and magnetic theories,
$\widetilde x_{D_{k+2}}^{min}\neq x^{min}_{D_{k+2}}$, with the magnetic
value of $\widetilde x^{min}_{D_{k+2}}$ somewhat smaller; they are a
little different
because the magnetic theory contains the extra singlet fields $M_{\ell
j}$
and the additional superpotential terms.
\bigskip
\centerline{\epsfxsize=0.50\hsize\epsfbox{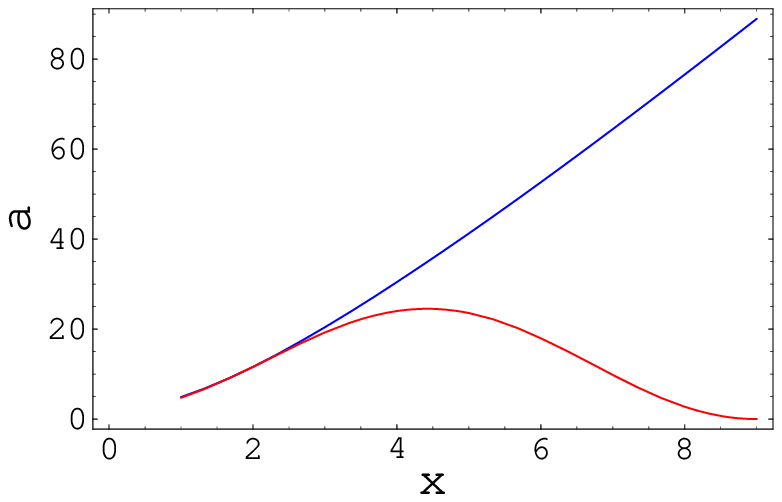}}
\centerline{\ninepoint\sl \baselineskip=8pt {\bf Figure 17:}
{\sl The central charge for irrelevant (top, blue) and relevant 
(bottom, red)
$\widetilde{X}^4$.}}
\bigskip

We can get an upper bound on  $\widetilde x_{D_{k+2}}^{min}$ as in
\xdkminii, by using the fact that the $\widetilde y(\widetilde x)$
obtained by the
above procedure is a
monotonically decreasing function of $\widetilde x$.
This  implies
\eqn\magxdkmin{{\widetilde{x}^{min}_{D_{k+2}}\over k+1}<1-\widetilde
y_{asymp}\approx 1.1038.}  Here $\widetilde y_{asymp}$ is the value
that $\widetilde y$ plateaus at for a while, when $\widetilde x$ is
large
but still below $\widetilde x^{min}_{D_{k+2}}$.  We compute $\widetilde
y_{asymp}$ in Appendix D.
{}From \magxdkmin , we can then calculate  $\widetilde
x_{D_{k+2}}^{min}$ in
the large $k$ limit:
\eqn\magxlgk{\widetilde{x}^{min}_{D_{k+2}} \approx 1.1038 k.}
Again, we see that $\widetilde x^{min}_{D_{k+2}}$
\magxlgk\ of the magnetic theory is slightly smaller than that of the
corresponding
electric theory \xdklg.

Figure 18 shows, for the magnetic $D_5$ example,  the
R-charges $R[q]\equiv \widetilde y(\widetilde x)$ as computed by the
above a-maximization procedure.  Though it was obtained by patching
together
the $\widetilde y^{(p)}(\widetilde x)$, it is continuous.  In
fig.
19 we plot $R[X]$ for this same theory, given as in \mrxysquared.  $\Tr
\widetilde X^4$ becomes relevant when $R(\widetilde X)\rightarrow
\half$, which is
found from fig. 19 to occur for $\widetilde x>\widetilde
x^{min}_{D_5}\approx
1.86$.

\bigskip
\centerline{\epsfxsize=0.50\hsize\epsfbox{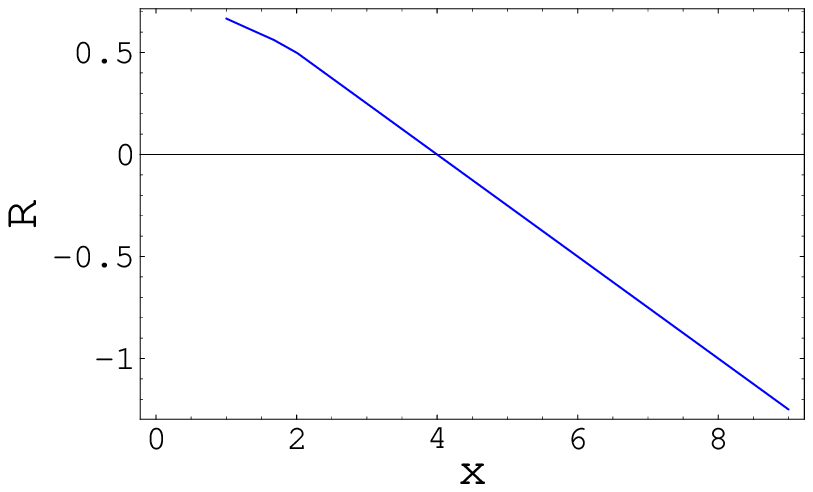}}
\centerline{\ninepoint\sl \baselineskip=8pt {\bf Figure 18:}
{\sl $\widetilde{y}(\widetilde{x})=R(q)$ for magnetic $D_5$.}}
\bigskip

\bigskip
\centerline{\epsfxsize=0.50\hsize\epsfbox{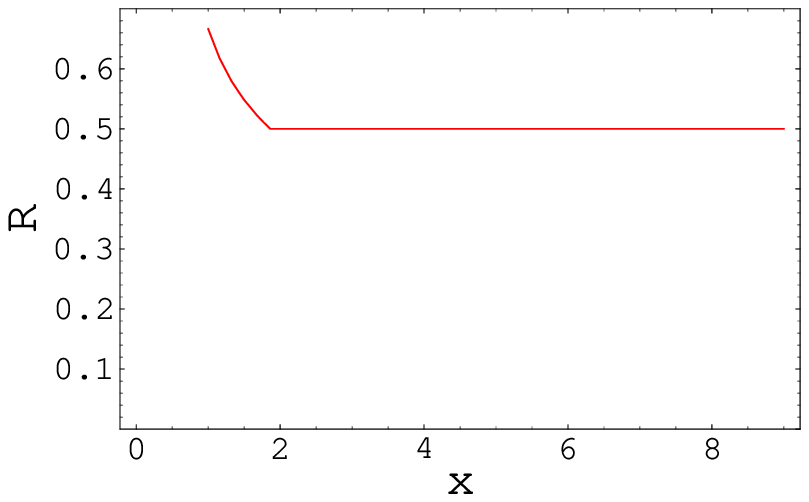}}
\centerline{\ninepoint\sl \baselineskip=8pt {\bf Figure 19:}
{\sl $R(\widetilde{X})$ for magnetic $D_5$.}}
\bigskip

We see (as in the $A_k$ case discussed in \KPS) that there is
a conformal window in which the $D_{k+2}$ superpotential $\Tr (X^{k+2}+
XY^2)$ and its magnetic analog are both relevant:
\eqn\emaf{x^{min}_{D_{k+2}} < x <3k-\widetilde{x}^{min}_{D_{k+2}}.}
For example, for $k=3$, $x_{D_5}^{min}\approx 2.09$
and $\widetilde{x}_{D_5}^{min} \approx
1.86$.
For the case of large $k$, we can use our results \xdklg\ and
\magxlgk\ to show that this window always exists:
\eqn\largekwin{{9 \over 8}k=x^{min}_{D_{k+2}} < x
<3k-\widetilde{x}^{min}_{D_{k+2}} \approx 2.062k.}
Within this window the central charges agree, as they should:
\eqn\ccagree{a^{el}(x)=\widetilde{a}^{m}(3k-x).}
Outside the window, the charges do not agree.  Indeed they should
not have been expected to agree, because one or the other side
does not readily exhibit the accidental symmetries of the IR theory.
In fig. 20, we have plotted the difference between the
(numerically computed) electric and magnetic central charges for the
$D_5$ theory and its dual. As one can see, they agree
in the range $2.09<x<9-1.86=7.14$ but disagree outside.
The correct $a$ to use is the larger of $a^{el}$ or $\widetilde
a^{mag}$, and it's
larger because of maximizing $a$ over a bigger symmetry group.

\bigskip
\centerline{\epsfxsize=0.50\hsize\epsfbox{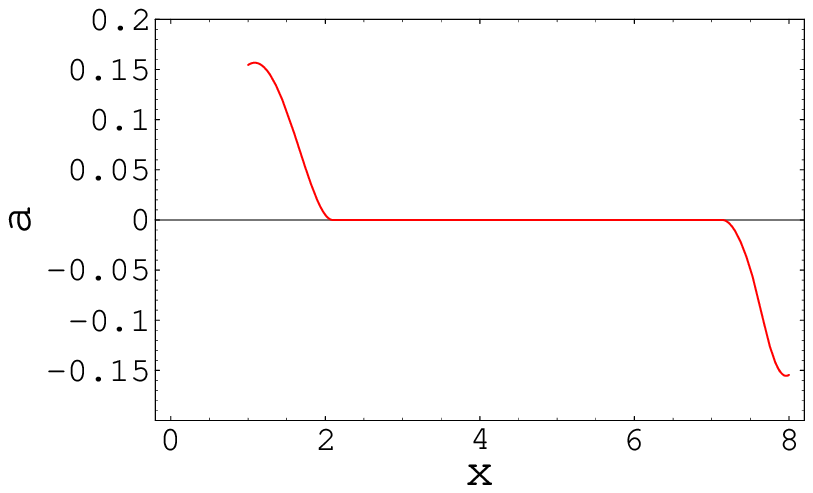}}
\centerline{\ninepoint\sl \baselineskip=8pt {\bf Figure 20:}
{\sl $a^{el}_{D_5}(x)/N_f^2 - \widetilde{a}^{mag}_{D_5}(9-x)/N_f^2$.}}
\bigskip

\newsec{The $E_{6,7,8}$ RG fixed point theories}

\subsec{Chiral rings and stability bounds}
Before discussing the $E_{6,7,8}$ RG fixed points in detail, we make
some
general comments about their chiral rings and the possible existence of
a
stability bound.   We'll consider the $E_6$ example; the $E_7$ and
$E_8$
cases are similar.

The superpotential $W_{E_6}=\Tr (Y^3+X^4)$ leads to the chiral ring
relations
\eqn\eomvi{\partial _XW_{E_6}=X^3=0, \qquad \partial _YW_{E_6}=Y^2=0.}
(For simplicity we consider $U(N)$ to avoid having to impose
tracelessness).    These do not truncate the chiral ring; e.g. they do
not allow us to eliminate $(XY)^n$ for arbitrary $n$.

The stability bound is related to the counting of vacua upon
deformation,
which is related to the number of independent products in the chiral
ring.  Roughly $x^{max}$ equals the number of independent products of
$X$ and $Y$ in the chiral ring.  To be more precise, upon deforming
the superpotential by the generic, lower order terms, the equations of
motion will have representations of various dimensions.  In the $A_k$
case, the chiral ring reps are all one dimensional, because we can
diagonalize the single adjoint. For the $D_{k+2}$ case we saw, following
\CKV, that the deformed ring relations \dkceom\ have $k+2$ different
one-dimensional
reps and $\half (k-1)$ different two-dimensional representations.

More generally, suppose that some deformed superpotential has $n_d$
different
$d$-dimensional representations, for $d=1,2,3\dots$.  We can take the
adjoints
to have $N_{d,i_d}$ copies of the $i_d$'th d-dimensional
representation, with $i_d=1
\dots n_d$, breaking the gauge group as
\eqn\genbrk{U(N_c)\rightarrow \prod _d\prod _{i_d=1}^{n_d}
U(N_{d,i_d}), \qquad \hbox{with}\quad N_c=\sum _d\sum
_{i_d=1}^{n_d}dN_{d,i_d}.}
The $U(N_{d,i_d})$ theory has $dN_f$ massless flavors and no massless
adjoints,
and will be stable if we can find a solution such that all
$N_{d,i_d}<dN_f$.  It
thus follows that
\eqn\genstabb{N_c<\sum _d n_d d^2N_f, \qquad \hbox{i.e.} \quad
x^{max}=\sum _{d=1}^\infty
n_d d^2.}

For the $A_k$, $D_{k+2}$, and $E_{6,7,8}$ Landau-Ginzburg
superpotentials, the
number of 1d representations is always the rank of the corresponding ADE
group, i.e. $n_{d=1}=r$.  In particular, for $E_6$, we have $n_1=6$,
which would be the dimension of the chiral ring if $X$ and $Y$ were not
matrices.  From \genstabb\
it follows that $x^{max}\geq r$, e.g. for $E_6$, $x^{max}_{E_6}\geq 6$.
   We can see that
if the chiral ring does not truncate, corresponding to having
arbitrarily many different
representations of the deformed superpotential EOM, then the sum in
\genstabb\
will be infinite and so $x^{max}=\infty$, i.e. no stability bound.
This is what our
classical analysis \eomvi\ suggests for the $E_6$ SCFT: an infinite
classical
result for $x^{max}_{E_6}$.

But perhaps the $E_6$ chiral ring truncates at the quantum
level, as we already suggested should be the case for the
even $k$ case of the $D_{k+2}$  chiral ring.  In this case,
there would be a quantum stability bound $x^{max}_{E_6}$
which is finite.  Our numerical analysis of this section, combined with
our
belief in the a-theorem, suggest that there is indeed a finite quantum
stability bound $x^{max}$ for the $E_6$, and $E_7$ and $E_8$ RG
fixed points (where, similar to the $E_6$ case, the classical ring
relations
do not suffice to truncate the ring).

Finally, as we discussed in \xmaxred, we expect that RG flows always
reduce
the stability bound $x_{IR}^{max}<x_{UV}^{max}$.  E.g. we can flow from
$E_6$
to $D_5$, so we expect $x^{max}_{E_6}>x^{max}_{D_5}=9$.

\subsec{$E_6$:  $W=\Tr (X^4+Y^3)$}

As seen in sect. 3.2, for $x>x_{E_6}^{min}\approx 2.55$,
there is a new RG fixed point associated with $W=\lambda _1\Tr X^4+
\lambda _2\Tr Y^3$.  Starting e.g. near the $\widehat O$ RG fixed point
and perturbing by this superpotential, the $\lambda _1$ term is
initially irrelevant while the $\lambda _2$ term is relevant, driving
the theory
near the $\widehat E$ RG fixed point.  Then, for $x>x_{E_6}^{min}$, the
$\lambda _1$ term becomes relevant and drives the theory to the new
$\widehat E_6$ RG fixed point.  The flow picture is analogous to fig.
13.
The superconformal $U(1)_R$ charges at the $E_6$ fixed point
are determined by the superpotential and the anomaly free condition
to be
\eqn\revi{R(Q)=R(\widetilde Q)=1-{x\over 6} , \quad R(X)={1\over 2},
\quad R(Y)={2\over 3}.}

We can compute the central charge $a(x)$, using the R-charges \revi,
correcting it to account for the gauge invariant operators which
hit the unitarity bound and then become free fields.  This procedure is
straightforward (but tedious).  In addition to the mesons which hit
the unitarity bound, we will argue in Appendix C that baryons would
also hit the unitarity bound if we continued to $x$ sufficiently large.
   For
the present discussion it suffices to take $x\leq 14$, for which one
can check that no baryon
has yet hit the unitarity bound.  So we only need to account for the
mesons which hit the bound, along with their multiplicities; e.g.
one must account for the fact the two mesons
$\widetilde{Q}XYQ$ and $\widetilde{Q}YXQ$ are distinct.
Accounting for all this, we numerically obtained the central charge
$a_{E_6}(x)$
for $x\leq 14$.

A prediction of the a-theorem is that $a$ should monotonically
decrease as a function of $N_f$, for fixed $N_c$.  As discussed
in sect. 2.3, this implies
that $a(x)x^{-2}N_f^{-2}$
must be a strictly decreasing function of $x$.  We plotted
$a_{E_6}(x)N_f^{-2}x^{-2}$
in fig. 21
\bigskip
\centerline{\epsfxsize=0.50\hsize\epsfbox{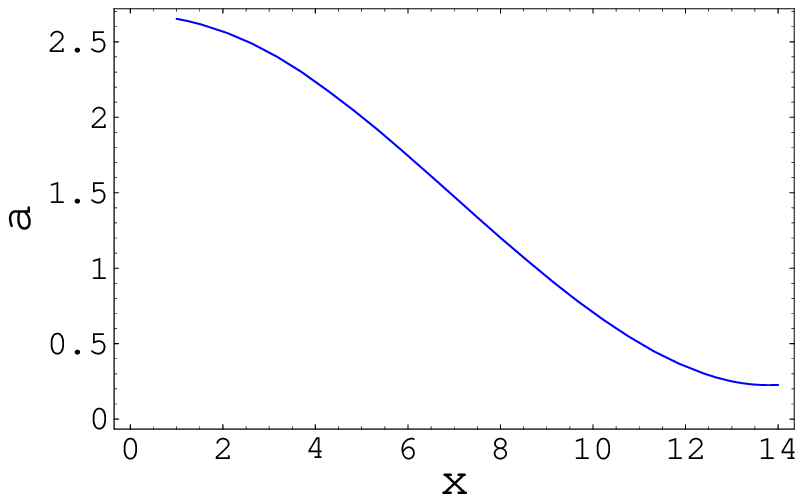}}
\centerline{\ninepoint\sl \baselineskip=8pt {\bf Figure 21:}
{\sl $a(x)x^{-2}N_f^{-2}$ for the $E_6$ theory.}}
\bigskip

\noindent
Contrary to the a-theorem prediction for a monotonically
decreasing function, the curve in fig. 21 flattens out at
$x\approx 13.80$.  There is a similar flattening out for the
$D_{k+2}$ theories right above the stability bound
$x^{max}_{D_{k+2}}=3k$.  Since we, by now, have much faith in the
a-conjecture, and we also suspected anyway that the $E_6$
RG fixed point might have a quantum truncated chiral ring
and stability bound, this is what we think the flattening in fig.
21 is showing: that there is indeed a quantum stability bound at
$x^{max}_{E_6}<13.80$ (which is, fortunately, consistent
with our earlier statements that $x^{max}_{E_6}>6$ and, via
the RG flow to $D_5$, $x^{max}_{E_6}>9$).
We leave a deeper
understanding of the quantum stability bounds such as $x^{max}_{E_6}$
as an open question for future work.

We verified the $a$-theorem
prediction that $a_{\widehat E}(x)>a_{E_6}(x)$ for all $x$ in the range
where the $E_6$ fixed point exists: $x>x_{E_6}^{min}
\approx 2.55$. One can see this behavior in fig. 22.

\bigskip
\centerline{\epsfxsize=0.50\hsize\epsfbox{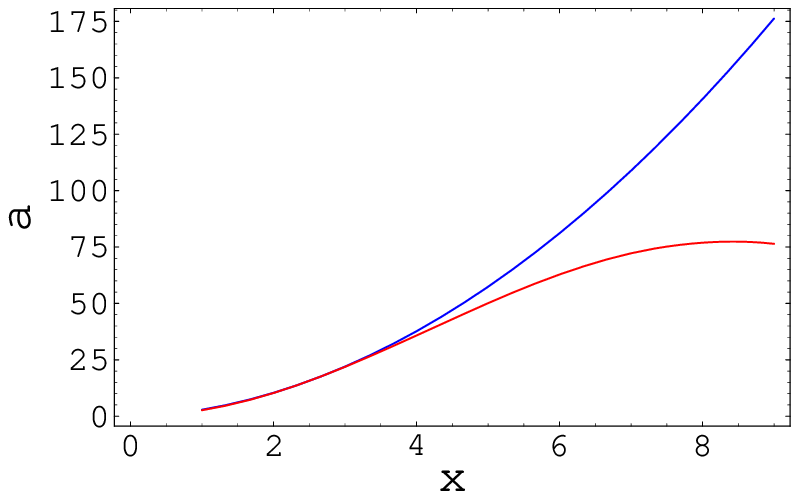}}
\centerline{\ninepoint\sl \baselineskip=8pt {\bf Figure 22:}
{\sl $a_{\widehat{E}}/N_f^2$ (top, blue) and $a_{E_6}/N_f^2$
(bottom, red).
The curves touch at $x_{E_6}^{min} \approx 2.55$.}}
\bigskip
\noindent
Because there is a relevant $\Delta W$ deformation taking
$E_6\rightarrow D_5$, the a-theorem predicts that
$a_{E_6}(x)>a_{D_5}(x)$ for the range of $x$, $x>x_{E_6}^{min}$ where
both
RG fixed points exist.   We have plotted $a/N_f^2$ for
both $D_5$ and $E_6$ in Figure 23. The a-theorem is
potentially violated in the region $x<1.50$, but is indeed satisfied
in the entire region $x>x_{E_6}^{min}\approx 2.55$, where the $E_6$ SCFT
actually exists.

\bigskip
\centerline{\epsfxsize=0.50\hsize\epsfbox{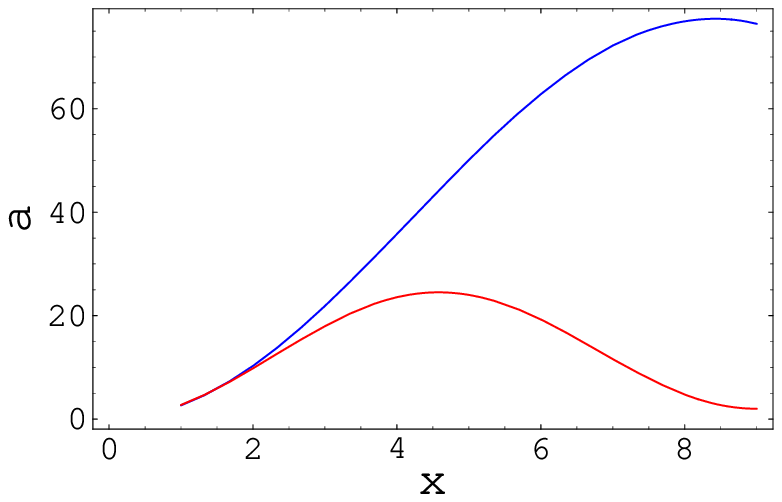}}
\centerline{\ninepoint\sl \baselineskip=8pt {\bf Figure 23:}
{\sl $a_{E_6}/N_f^2$ (top, blue) and $a_{D_5}/N_f^2$ (bottom, red).
The curves cross at $x\approx 1.50$.}}
\bigskip

\subsec{$E_7$: $W=\Tr (YX^3+Y^3)$}

Starting with $W=\lambda _1\Tr YX^3+\lambda _2\Tr Y^3$, we have a
RG flow that goes first to the vicinity
of the $\widehat E$ fixed point and then, provided that $x>x_{E_7}^{min}
\approx 4.12$, the $\lambda _1$ term takes over and
drives the theory to the new $E_7$ RG fixed point.  At the $E_7$ RG
fixed point the superconformal R-charges are determined to be
\eqn\revii{R(Q)=R(\widetilde Q)=1-{x\over 9}, \quad R(X)={4\over 9},
\quad R(Y)={2\over 3}.}

In computing the central charge $a_{E_7}(x)$, we must account for
all of the independent mesons and baryons which hit the unitarity bound.
Again, doing so is quite tedious, so we only carried out the analysis to
relatively low $x$.  In the range analyzed, we
   verified the a-theorem predictions.  For example
there is a RG flow $\widehat E\rightarrow E_7$, and we verified
the a-theorem prediction that  $a_{\widehat E}(x)>a_{E_7}(x)$
for all $x>x_{E_7}^{min}$; see fig. 24.

\bigskip
\centerline{\epsfxsize=0.50\hsize\epsfbox{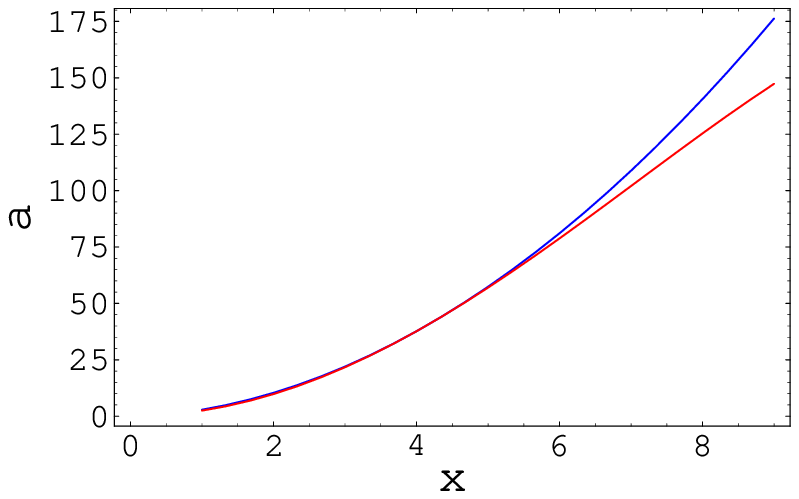}}
\centerline{\ninepoint\sl \baselineskip=8pt {\bf Figure 24:}
{\sl $a_{\widehat{E}}/N_f^2$ (top, blue) and $a_{E_7}/N_f^2$ (bottom,
red).
The curves touch at $x_{E_7}^{min} \approx 4.12$.}}
\bigskip

At the $E_7$ RG fixed point, the superpotential $\Delta W=\Tr X^4$
is relevant, since \revii\ gives $R(\Delta W)=
16/9<2$, and it leads to the RG flow $E_7\rightarrow E_6$.
The $a$ theorem thus requires that $a_{E_7}(x)>a_{E_6}(x)$ for
all $x > x_{E_7}^{min}$.  Looking at fig. 25, we see that this
inequality
is indeed satisfied. Again there is an apparent violation, since
$a_{E_6}$ is larger than $a_{E_7}$ for
$x<3.16$, but actually no contradiction with the a-theorem because
the $E_7$ fixed point does not exist for $x<x^{min}_{E_7}\approx 4.12$,
so there
is no a-theorem violating RG flow.

\bigskip
\centerline{\epsfxsize=0.50\hsize\epsfbox{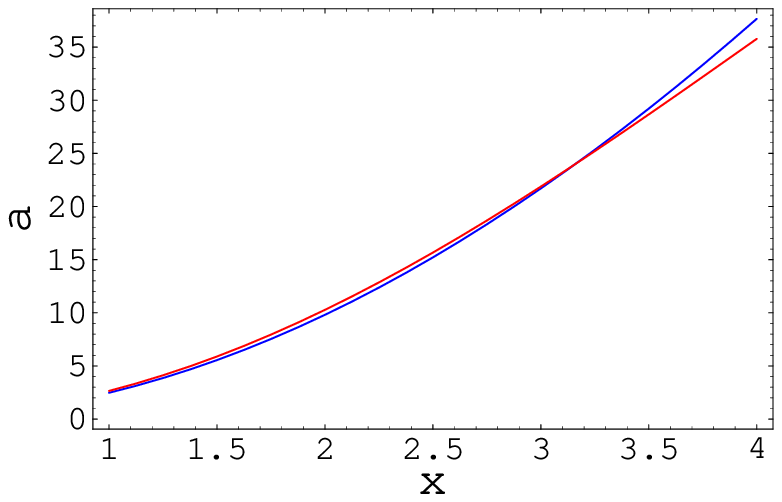}}
\centerline{\ninepoint\sl \baselineskip=8pt {\bf Figure 25:}
{\sl $a_{E_7}/N_f^2$ (blue) and $a_{E_6}/N_f^2$ (red).
The curves cross at $x \approx 3.16$.}}
\bigskip

As in the $E_6$ case discussed above, we suspect that there is
a stability bound upper limit $x_{E_7}^{max}$ for the $E_7$ fixed
points.  Since we can RG flow from $E_7\rightarrow E_6$, it should
satisfy $x^{max}_{E_7}>x^{max}_{E_6}$.

\subsec{$E_8$: $W=\Tr (X^5+Y^3)$}

Taking $W=\lambda _1\Tr X^5+\lambda _2\Tr Y^3$, there is an RG flow
first in $\lambda _2$ to the vicinity of the $\widehat E$ RG fixed
points, and then the $\lambda _1$ term drives the theory to the
new $E_8$ RG fixed points provided that $x>x_{E_8}^{min}\approx
7.28$.   At the $E_8$ superconformal fixed point, the R-chargers
are
\eqn\reviii{R(Q)=R(\widetilde Q)=1-{x\over 15}, \quad R(X)={2\over
5}, \quad R(Y)={2\over 3}.}  (We note from \rdkis, \revi, \revii,
\reviii\ that the $A_k$, $D_{k+2}$ and $E_{6,7,8}$ RG fixed points have
$R(Q)=R(\widetilde Q)=1-{2x\over h}$ where $h$ is the dual Coxeter
number of the corresponding $A_k$, $D_{k+2}$, or $E_{6,7,8}$ group.)

The central charge $a_{E_8}(x)$ can thus be computed,  where we again
must
account for the accidental symmetries associated with mesons and baryons
which hit the unitarity bound.  We carried out this process to $x=16$.
As seen in fig. 26,
the a-theorem
prediction $a_{\widehat E}(x)>a_{E_8}(x)$ is satisfied for
$x>x_{E_8}^{min}$,
where the $E_8$ fixed point exists.

\bigskip
\centerline{\epsfxsize=0.50\hsize\epsfbox{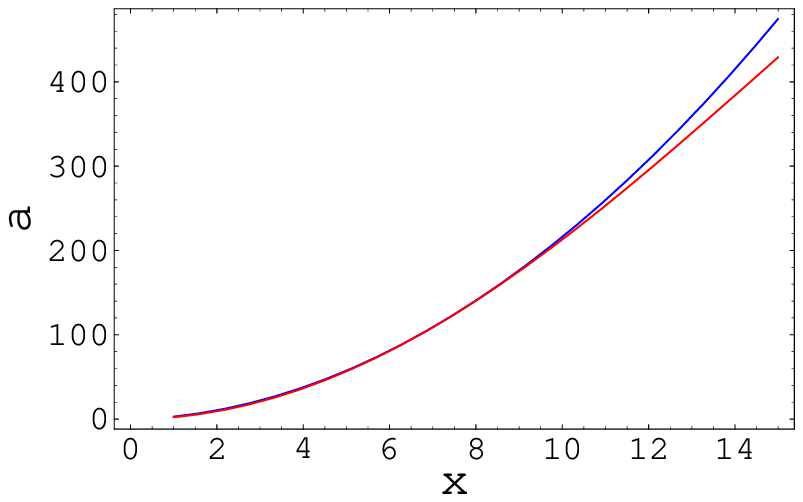}}
\centerline{\ninepoint\sl \baselineskip=8pt {\bf Figure 26:}
{\sl $a_{\widehat{E}}/N_f^2$ (top, blue) and $a_{E_8}/N_f^2$
(bottom, red).
The curves touch at $x_{E_8}^{min} \approx 7.28$.}}
\bigskip

Starting from the $E_8$ fixed point, the deformation
$\Delta W=\Tr YX^3$ is relevant, leading to the RG flow
$E_8\rightarrow E_7$.  We can also deform the $E_8$ fixed
point by the relevant deformation  $\Delta W=\Tr X^4$, which leads to
the RG flow $E_8\rightarrow E_6$.
So the a-theorem predicts that $a_{E_8}(x)>a_{E_7}(x)$ and
$a_{E_8}(x)>a_{E_6}(x)$, for all $x$ in the range where the $E_8$ fixed
point
exists, $x>x_{E_8}^{min}$.  We see from fig. 27 that the first of
these
inequalities is indeed satisfied; the second works similarly.
    The inequality $a_{E_8}(x)>a_{E_7}(x)$ is satisfied
for $x>5.25$ and the inequality $a_{E_8}(x)>a_{E_6}(x)$ is satisfied for
$x>3.79$; so both are satisfied for $x>x_{E_8}^{min}\approx 7.28$.

\bigskip
\centerline{\epsfxsize=0.50\hsize\epsfbox{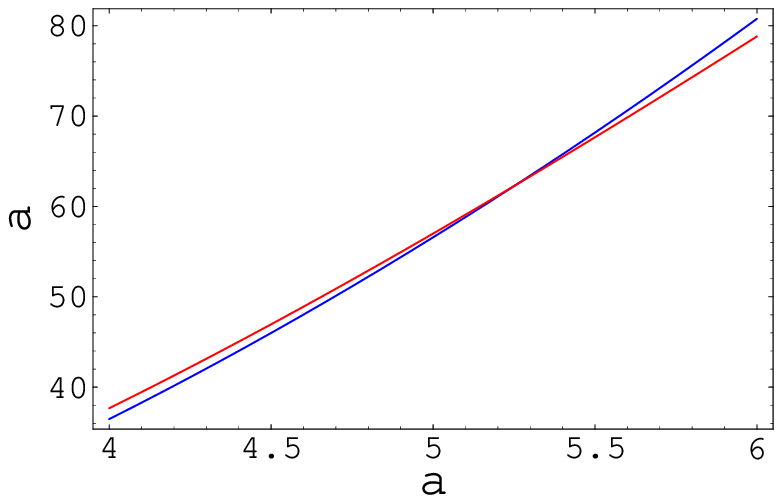}}
\centerline{\ninepoint\sl \baselineskip=8pt {\bf Figure 27:}
{\sl $a_{E_8}/N_f^2$ (blue) and $a_{E_7}/N_f^2$ (red).
The curves cross at $x \approx 5.25$.}}
\bigskip

As in the $E_6$ and $E_7$ cases discussed above, we suspect that there
is
a quantum stability bound upper limit $x_{E_8}^{max}$ for the $E_8$
fixed
points.  It should satisfy $x_{E_8}^{max}>x_{D_7}^{max}=15$ since we
can RG
flow from $E_8\rightarrow D_7$.

\newsec{New RG fixed points, with mesonic superpotentials}

We briefly mention some new SCFTs, which can be obtained from our
previously discussed ones by relevant superpotential deformations
involving the meson chiral operators.

\subsec{Flowing from $\widehat O$}

Starting at the $\widehat O$ SCFT, the mesonic  gauge
invariant operators \mesgen, with $n$ powers of
$X$ or $Y$, will be relevant if
$2y(x)+nz(x)<2$.  Using the fact that the expressions \maxy\ and \maxz\
for $y(x)$
and $z(x)$ are monotonically decreasing with $x$, with
asymptotic values \ohatrlim, we see that we must have $2(0.575)+\half
n<2$. Thus, the only solution is $n=1$, i.e. the superpotential
\eqn\ohatmesd{\Delta W=\lambda \widetilde Q_iYQ^i,}
where we chose to break $SU(N_f)\times SU(N_f)\rightarrow SU(N_f)$ by
including all meson flavor components diagonally.  The superpotential
\ohatmesd\
is relevant for all $x\geq 1$.  This relevant deformation drives
$\widehat O$ to a
new RG fixed point, which we'll call $\widehat O_M$.  We expect it to
flow to an interacting SCFT by the argument outlined in footnote 5.

At the $\widehat O_M$ SCFT, the R-charges are given by the general
expression
\rtwoadj\ with the additional constraint $2y+z=2$.   Using this to
eliminate $z$,
the a-maximization procedure can be used to solve for $y(x)$.   We
leave analysis
of this to the interested reader.

\subsec{Flowing from $\widehat D$}

Starting at the $\widehat D$ RG fixed point, we can see from the
asymptotic
values of the R-charges \yasymp\ that all of the mesons \dhatmes\ can be
relevant operators:
\eqn\dhatmesii{\Delta W _{M_{\ell j}}=\widetilde Q_iX^\ell Y^jQ^i,
\qquad
\ell \geq 0, \quad j=0,1,}
will be relevant for any $\ell$ and $j=0,1$, provided that $x$ is
sufficiently large,
$x>x^{min}_{\widehat D_{M_{\ell j}}}$.
The interested reader can analyze the resulting RG
fixed point theories, where the superconformal R-charge is completely
fixed
by the condition that it respect the two terms in the superpotential.

\subsec{Flowing from $\widehat E$}

Using the asymptotic values in \relim\ for $R(Q)$ and $R(X)$, together
with
$R(Y)=2/3$, we see that the relevant mesonic deformations are
\eqn\ehatmeson{\Delta W=\widetilde Q_i X^\ell Q^i, \quad\hbox{for}\quad
0<\ell \leq 3, \quad\hbox{and} \quad \Delta W=\widetilde Q_i XY Q^i.}
Again, the interested reader can analyze the resulting
RG fixed point theories, where the superconformal R-charge is completely
fixed by the condition that it respect the two terms in the
superpotential.

\newsec{New SCFTs, with $\widehat O$ not in their domain of attraction?}

Our process of starting at $\widehat O$ and then following RG flows to
new SCFTs, as in fig. 1, could potentially miss SCFTs, if they do
not have $\widehat O$ in their domain of attraction.  As a concrete
example
we ask if there might be $W=\Tr Y^{k+1}$ RG fixed points for $k>2$.

As seen sect. 2, $W=\lambda \Tr Y^3$ is a
relevant deformation of the $\widehat O$ RG fixed point, driving the
$\widehat O$ RG fixed points to the $\widehat E$ RG fixed points
for all $N_f$.  We have also seen that $W=\Tr Y^{k+1}$ for $k>2$ is
an irrelevant deformation of the $\widehat O$ RG fixed point. That is,
if
we add $W=\lambda \Tr Y^{k+1}$ for $k>2$, then $\beta (\lambda )>0$,
at least for small $\lambda$, and $\lambda \rightarrow 0$ in the IR.
We might wonder, though, if $\beta (\lambda)$ could perhaps look
like that of fig. 28, and nevertheless have a zero at some critical
value
$\lambda _*$ corresponding to some new hypothetical  RG fixed points,
which
we name $G_k$, which do not have
$\widehat O$ in their domain of attraction.

\bigskip
\centerline{\epsfxsize=0.50\hsize\epsfbox{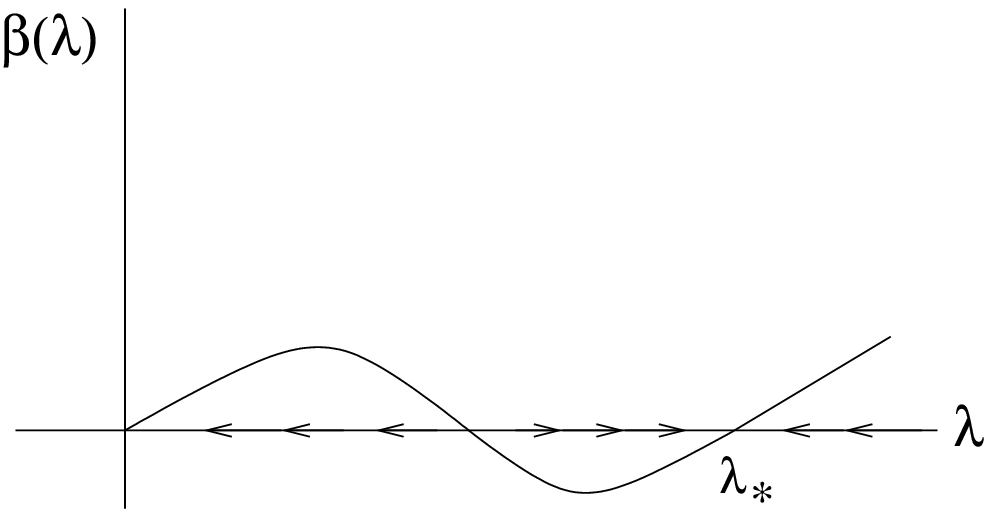}}
\centerline{\ninepoint\sl \baselineskip=8pt {\bf Figure 28:}
{\sl A $G_k$ fixed point?}}
\bigskip

Our personal prejudice is that such a hypothetical new RG fixed point,
requiring large Yukawa coupling as in fig. 28, seems unlikely.  But
we'll
here take an unbiased view and see what constraints could be placed on
such
hypothetical additional RG fixed points.

Assuming that the superpotential $W=\Tr Y^{k+1}$ does control some
hypothetical $G_k$ RG fixed point,
the anomaly free R-symmetry with $R(W)=2$ is the one
parameter family
\eqn\rwithw{R(Y)={2 \over k+1}, \qquad R(Q)=R(\widetilde{Q})\equiv
y, \qquad
R(X) = {1+x-y \over x} - {2 \over k+1}.}
We can now maximize the central charge $a$, given by \atwoadj\ with
$z=2/(k+1)$, with respect to $y$; the result is
\eqn\ywonek{y=1+{x(6-\sqrt{-1+74x^2+k^2(-1+2x^2)+k(-2+4x^2)}) \over
3(1+k)(-1+2x^2)}.}
For all $k$, this $y$ is such that no gauge invariant chiral operator
ever violates the unitarity bound $R\geq 2/3$ (since $y>1/3$ for all
$x$, if $k\geq 2$).  So unitarity does not require the existence of
accidental symmetries,
and the above results could be correct as they stand, without
modification.  We can now plug \ywonek\ back into \atwoadj\ to get
$a_{G_k}(x)$.

We now consider the deformation of the hypothetical $G_k$ fixed
points by $\Delta W=\Tr YX^2$ which, if relevant, will lead to RG flows
either as $G_k\rightarrow \widehat D$, or as $G_k\rightarrow D_{k+2}$,
depending on whether $\Delta W$ wins over $\Tr Y^{k+1}$, or if
both remain important in the IR, respectively.
To determine when $\Delta W$ is relevant, consider its
superconformal R-charge:
\eqn\rwgk{R(YX^2) = 2 \left ({1 - y \over x} +1 \right ) -
{2 \over k+1}.}
For $x$ close to 1, we see from \ywonek\ that $R(YX^2)>2$ and the
deformation is irrelevant.  On the other hand, for $x\rightarrow
\infty$,
we get $R(YX^2)\rightarrow  2k/(k+1)<2$ and $\Delta W$ is relevant,
so there is some critical value of $x$ where this deformation
becomes relevant.   We
can now check if the hypothetical $G_k\rightarrow \widehat D$ and 
$G_k\rightarrow
D_{k+2}$ RG
flows are compatible with the conjectured a-theorem, which would require
$a_{G_k}(x)>a_{\widehat D}(x)$, or $a_{G_k}(x)>a_{D_{k+2}}(x)$ for all
$x$ such that the RG fixed points exist.

We did some checks of this, hoping to use the conjectured a-theorem to 
rule
out the hypothetical $G_k$ SCFTs, or at least place some stringent 
bounds
on them.  What we found, however, is that all potential a-theorem 
violations
only occur for low values of $x$.  So, assuming the a-theorem, we can 
at best
rule out the hypothetical $G_k$ fixed points for low $x$, but we were 
unable
to rule them out entirely.

We could, similarly, consider the possibility that our
  $A_k$,  $D_{k+2}$ or $E_{6,7,8}$
RG fixed points exist for values of $x$ below the $x^{min}$ found for 
$A_k$ in
\KPS\ and $D_{k+2}$ and $E_{6,7,8}$ here.  All of these values of
$x^{min}$ were based on having $\widehat O$ in the domain of 
attraction.  But
perhaps, for example, the $D_{k+2}$ RG fixed points could actually also 
exist
for $x$ in some range $x^{mini}_{D_{k+2}}<x<x^{min}_{D_{k+2}}$, which 
is outside of
$\widehat O$'s domain of attraction.   If the hypothetical new lower
bound for existence of the  $D_{k+2}$ SCFT,  $x^{mini}_{D_{k+2}}$, 
extended too far below
the values for $x^{min}_{D_{k+2}}$ found in sect. 4, we would have
$D_{k+2}
\rightarrow D_{k'+2}$ RG flows violating the a-theorem conjecture, as
seen e.g. in fig. 16 if $x_{D_6}^{mini}<2.56$.  Given our belief in the
validity of the
a-theorem, we expect that any hypothetical $D_{6}$
RG fixed points outside of $\widehat O$'s domain of attraction, for 
example, could
only exist for the $x$ range $2.56\leq 
x^{mini}_{D_6}<x<x^{min}_{D_6}\approx 3.14$.

\newsec{Comments}

For the a-maximization method to be useful, we must know the full
symmetry
group of the IR fixed point, including all accidental symmetries.  The
class
of such accidental symmetries having to do with operators hitting the
unitarity bound and becoming free are easy to spot and
account for, as in \KPS\ and the present paper.  But  there can be other
accidental symmetries, which (to quote \KPS) did not have such an
obvious ``smoking gun"
characteristic as apparent unitarity violation.    For example, in SQCD,
the unitarity bound would
tell us that the meson becomes free for $N_f\leq {3\over 2}N_c$.  But
we know
{}from Seiberg duality \NSd\ that that's not the whole story: the
entire theory is in a free magnetic phase for $N_f\leq {3\over 2}N_c$.
   It would be nice if there were some
other tractable smoking gun tests which would reveal, even to someone
who did not know
about the Seiberg dual, that free mesons alone do not suffice for
$N_f\leq {3\over 2}N_c$.

Likewise, in sect. 6, we saw that the
$D_{k+2}$ conformal window does not extend above
$x>3k-\widetilde x_{D_{k+2}}^{min}$; knowing the dual \JB, we found that
the IR theory is then governed by the magnetic version of the $\widehat
D$.
Because we do not know the duals of any of our other RG fixed points
\oade, we can
not exclude the possibility that they too might have some new
symmetries at
sufficiently large $x$ (stronger coupling), which could be very obscure
in the original
Lagrangian descriptions -- but perhaps obvious in some as-yet-unknown
dual
descriptions.   Short of knowing such dual descriptions, it would be
nice if there
were a tractable test to alert us to the possible presence of such
accidental
symmetries, which go beyond those required by the unitarity bound.

\lref\MartinecZU{
E.~J.~Martinec,
``Algebraic Geometry And Effective Lagrangians,''
Phys.\ Lett.\ B {\bf 217}, 431 (1989).
}

\lref\VafaUU{
C.~Vafa and N.~P.~Warner,
``Catastrophes And The Classification Of Conformal Theories,''
Phys.\ Lett.\ B {\bf 218}, 51 (1989).
}
\lref\KachruAN{
S.~Kachru, S.~Katz, A.~E.~Lawrence and J.~McGreevy,
``Mirror symmetry for open strings,''
Phys.\ Rev.\ D {\bf 62}, 126005 (2000)
[arXiv:hep-th/0006047].
}
\lref\CachazoJY{
F.~Cachazo, K.~A.~Intriligator and C.~Vafa,
``A large N duality via a geometric transition,''
Nucl.\ Phys.\ B {\bf 603}, 3 (2001)
[arXiv:hep-th/0103067].
}

\lref\KatzAB{
S.~Katz,
``Versal deformations and superpotentials for rational curves in smooth
   threefolds,''
arXiv:math.ag/0010289.
}

\lref\CachazoSG{
F.~Cachazo, B.~Fiol, K.~A.~Intriligator, S.~Katz and C.~Vafa,
``A geometric unification of dualities,''
Nucl.\ Phys.\ B {\bf 628}, 3 (2002)
[arXiv:hep-th/0110028].
}

It would also be nice to connect the occurrence of Arnold's ADE
singularities
found here to other ways in which the ADE series arises in physics and
mathematics.
E.g. in 2d $\N =2$ theories the ADE superpotentials were special because
they led to the $\widehat c<1$ minimal models \refs{
\MartinecZU, \VafaUU}, which could be characterized by the requirement
that
all elements of the chiral ring be relevant.  Another occurrence of the
ADE
polynomials is via geometry, e.g. the ALE singularities and
generalizations,
which could connect with our SCFTs via string-engineering our SCFTs via
D-branes at the singularities.  It is indeed
possible to string-engineer some variants of our 4d $A_k$ and $D_{k+2}$
SCFTs (as well as non-conformal variants) via 4d spacetime filling D3
branes
at points (plus D5's wrapped on cycles for the non-conformal variants)
in a
suitable local Calabi-Yau geometry
\refs{\KachruAN, \CachazoJY, \KatzAB, \CKV, \CachazoSG}.  These
constructions do not
yield precisely our SCFTs, but rather these theories deformed by the
superpotential
$\Delta W=\widetilde Q_i XQ^i$ or $\Delta W=\widetilde Q_i Y Q^i$;
these are present
in the string constructions because they are based on a deformation
from $\N =2$.
\lref\DijkgraafDH{
R.~Dijkgraaf and C.~Vafa,
``A perturbative window into non-perturbative physics,''
arXiv:hep-th/0208048.
}

\lref\CachazoRY{
F.~Cachazo, M.~R.~Douglas, N.~Seiberg and E.~Witten,
``Chiral rings and anomalies in supersymmetric gauge theory,''
JHEP {\bf 0212}, 071 (2002)
[arXiv:hep-th/0211170].
}
\lref\DijkgraafXD{
R.~Dijkgraaf, M.~T.~Grisaru, C.~S.~Lam, C.~Vafa and D.~Zanon,
``Perturbative computation of glueball superpotentials,''
arXiv:hep-th/0211017.
}

It would also be interesting to connect our results about 4d SCFTs to
properties of these
theories upon breaking conformal invariance by generic relevant
superpotential
deformations, where some new techniques are available for analyzing the
effective glueball superpotentials and properties of the chiral ring
e.g. \refs{\CachazoJY,
\DijkgraafDH, \DijkgraafXD, \CachazoRY}.
   Perhaps there is some generalization of some of the fascinating
results in 2d $\N =2$ theories connecting properties
of SCFTs, such as the Poincare polynomial of superconformal R-charges,
and
properties of the critical points and solitons upon massive
deformations,
see e.g.
\ref\CecottiRM{
S.~Cecotti and C.~Vafa,
``On classification of N=2 supersymmetric theories,''
Commun.\ Math.\ Phys.\  {\bf 158}, 569 (1993)
[arXiv:hep-th/9211097].
}.

\centerline{\bf Acknowledgments}

We would like to thank Ben Grinstein, Kieran Holland, and Ronen Plesser 
for discussions.
This work was supported by DOE-FG03-97ER40546.

\appendix{A}{Some of the RG fixed points and flows, in the
perturbative regime}

We can now study the $\widehat O$, $\widehat D$ and $\widehat E$
RG fixed points of \oade\ in a perturbative regime, along the
lines of \refs{\GW, \BZ}. To do this, we take $SU(N_c)$ with $N_c$
large
and $N_f$ such that the theory is just barely asymptotically free.
Defining $x\equiv N_c/N_f$, the asymptotic freedom bound for the
theory with $N_a=2$ adjoint chiral matter fields is $x>1$.
The perturbative regime is $x=1+\epsilon$, with $0<\epsilon \ll 1$;
the RG fixed point couplings will be of order $\epsilon$, and we will
here
work only to leading order in $\epsilon$.  These results should be
qualitatively accurate provided that we tune $\epsilon$ to be
sufficiently
small by our choice of $N_c$ and $N_f$.

\subsec{The $\widehat O$ RG fixed points: $W=0$.}

In the limit $x=1+\epsilon$, with $0\leq \epsilon \ll 1$,
the gauge coupling beta function
$\beta (g)$ has a negative one loop part and a positive two loop
part. This leads as in \BZ\ to a perturbatively accessible RG fixed
point, with $\beta (g_*)=0$ solved by $g_*$ small.  A general expression
for the RG fixed point coupling of SUSY gauge theories in this limit is
\IW\
\eqn\gstarg{{g_*^2|{\cal G}|\over 8\pi ^2}\approx (6h-\sum _i n_i \mu
({\bf r}_i))
\left(\sum _j n_j \mu ({\bf r}_j)^2|{\bf r}_j|^{-1}\right)^{-1},}
with $|{\cal G}|$ the dimension of the gauge group, $2h$ the quadratic
Casimir
index of the adjoint representation, $n_i$ the number of matter chiral
superfields
in representation ${\bf r}_i$, and $\mu ({\bf r}_j)$ the quadratic
Casimir index
of ${\bf r}_j$.  For our present case this yields
\eqn\gstar{{g_*^2N_c\over 8\pi ^2}\approx {x-1\over 1+4x}\approx
{\epsilon \over 4}.}
We can thus make the 't Hooft coupling arbitrarily small by our choice
of $\epsilon$.  Because the fixed point coupling is small
for sufficiently small $\epsilon$,
we expect that perturbation theory is a reliable indicator of the
phase of the gauge theory  and that the theory is indeed in the
``non-Abelian Coulomb phase,'' which is an interacting conformal
field theory.

\subsec{The $\widehat E$ RG fixed points: $W={1\over 6}\lambda \Tr
Y^3$.}

In our perturbative regime
$x=1+\epsilon$, with $0\leq \epsilon \ll 1$,
the superpotential deformation does not affect
the gauge coupling beta function to leading order in $\epsilon$
(i.e. to one-loop, but there are higher loop effects).
So the gauge coupling remains
at the same fixed point value as in the $\widehat O$ case
\gstar.  The beta function for $\lambda$ is $\beta (\lambda)=
{3\over 2}\lambda \gamma (Y)$, which we can compute to leading order in
$\epsilon$. The one-loop anomalous dimension is
\eqn\gammalam{\gamma ( \lambda)=-{g_*^2N_c \over 8\pi^2}
+{\lambda^2 d^2 \over 32 \pi^2}.}
The $\lambda \Tr Y^3$ interaction vertex is $\lambda d^{abc}$
with $d^{abc}$ the
cubic Casimir
$d^{abc}\equiv \half \Tr[T^a \{ T^b, T^c \}]$ (with $a$ in the adjoint,
and $T^a$
in the fundamental representation of $SU(N_c))$.   It is
straightforward to show
that
\eqn\dsquared{d^{abc}d^{bce}={(N_c^2 - 4) \over
4N_c}\delta^{ae}\equiv d^2 \delta^{ae} .}
Plugging \gammalam\ and \gstar\ into the $\beta$ function
yields
\eqn\betalam{\beta (\lambda )={3\over 2}\lambda (-{g_*^2N_c \over
8\pi^2}+
{\lambda^2 d^2 \over 32 \pi^2})={3\over 2}\lambda ( {\lambda ^2 d^2
\over 32
\pi^2}
-{\epsilon \over 4}).}
The negative sign of \betalam\ for small non-zero $\lambda$ shows that
$\Tr Y^3$ is a relevant deformation of the $\widehat O$ RG fixed
point theory, which drives the theory to a new RG fixed point.
At this new RG fixed point, which we name $\widehat E$, we can find
the RG fixed
point coupling for the superpotential interaction:
\eqn\lamstar{\lambda _*^2={8 \pi^2 \epsilon \over d^2}}
to leading order in the $\epsilon \ll 1$ expansion.

\subsec{The $\widehat D$ RG fixed point theories: $W=\half \lambda \Tr
XY^2$}

Again, the added superpotential interaction $W=\half \lambda \Tr XY^2$
does
not affect the gauge coupling beta function to leading order in the
$\epsilon$ expansion.  So to leading order in $\epsilon$ the RG fixed
point coupling stays at the same value as in the $\widehat O$ and
$\widehat E$ cases, \gstar.  The beta function for $\lambda$ is $\beta
(\lambda )=\lambda (\half \gamma (X)+\gamma (Y))$, which we can evaluate
to be
\eqn\dbeta{\beta (\lambda )=\lambda (-3{g_*^2N_c\over 16\pi ^2}+{5 d^2
\over 64 \pi^2}
\lambda ^2)=\lambda (-{3\epsilon \over 8}+{5 d^2 \over 64 \pi^2 }
\lambda
^2).}
Again, this is negative at $\lambda =0$, showing that $W=\lambda \Tr
XY^2$
is a relevant perturbation of the $\widehat O$ RG fixed point, which
drives the theory to a new RG fixed point.  At that new RG gixed point,
which we name $\widehat D$, the superpotential interaction has the
fixed point value
\eqn\dlamfp{\lambda _*^2={24\pi^2 \epsilon \over 5 d^2}}
to leading order in the $\epsilon$ expansion.

\subsec{Combining the $\widehat D$ and $\widehat E$ interactions: the
$D_4$ SCFT}
Consider the theory with superpotential \eqn\dehat{W={1\over 6}\lambda
_1 \Tr
X^3+{1 \over 2} \lambda _2 \Tr XY^2.}  We have seen that either $\lambda
_1$
or $\lambda _2$ zero, with the other non-zero, drives the $\widehat
O$ RG fixed points to the $\widehat D$ and $\widehat E$ RG fixed points,
respectively.  We now consider the situation where both $\lambda _1$ and
$\lambda _2$ are taken to be non-zero.  Depending on the initial values
of $\lambda _1$ and $\lambda _2$, we could study these flows by e.g.
starting in the vicinity of the $\widehat O$, $\widehat D$, or $\widehat
E$ RG fixed points.  From all of these initial RG fixed points,
the added perturbation of \dehat\ is seen to be a relevant deformation.
We expect that $\widehat O$, $\widehat D$, and $\widehat E$ are all
in the same basin of attraction for a theory with both terms in the
superpotential
\dehat\ important; we name this new RG fixed point SCFT
$D_4$.

We can quantify and confirm
this picture in the limit where $x=1+\epsilon$ with
$0<\epsilon \ll 1$, where perturbation theory is valid.  Again,
to leading order, the superpotential \dehat\ does not affect the gauge
beta
so the gauge coupling fixed point value remains at the value \gstar.
The interesting flow is in the
couplings $\lambda _1$ and $\lambda _2$.
The beta functions for $\lambda _1$ and $\lambda _2$ are given by
\eqn\lambeta{\beta (\lambda _1)={3\over 2}\lambda _1 \gamma (X), \qquad
\beta (\lambda _2)=\lambda _2(\half \gamma (X) + \gamma (Y)).}
Computing the anomalous dimensions to one loop yields
\eqn\xyanomd{\eqalign{\beta (\lambda _1)&=\lambda _1(-3{g_*^2N_c\over
16\pi ^2}+
{3d^2 \over 64 \pi^2}(
\lambda _1^2+\lambda _2^2))\cr
\beta (\lambda _2)&=\lambda _2(-3{g_*^2N_c\over 16\pi ^2}+{d^2 \over 64
\pi^2}
(\lambda _1^2+5\lambda _2^2)).}}
These lead to flows which are attracted to the RG fixed point, at
\eqn\zerob{\lambda _{*1}^2=\lambda_{*2}^2={4\pi ^2\epsilon \over d^2}.}
as sketched in fig. 29.
\bigskip
\centerline{\epsfxsize=0.50\hsize\epsfbox{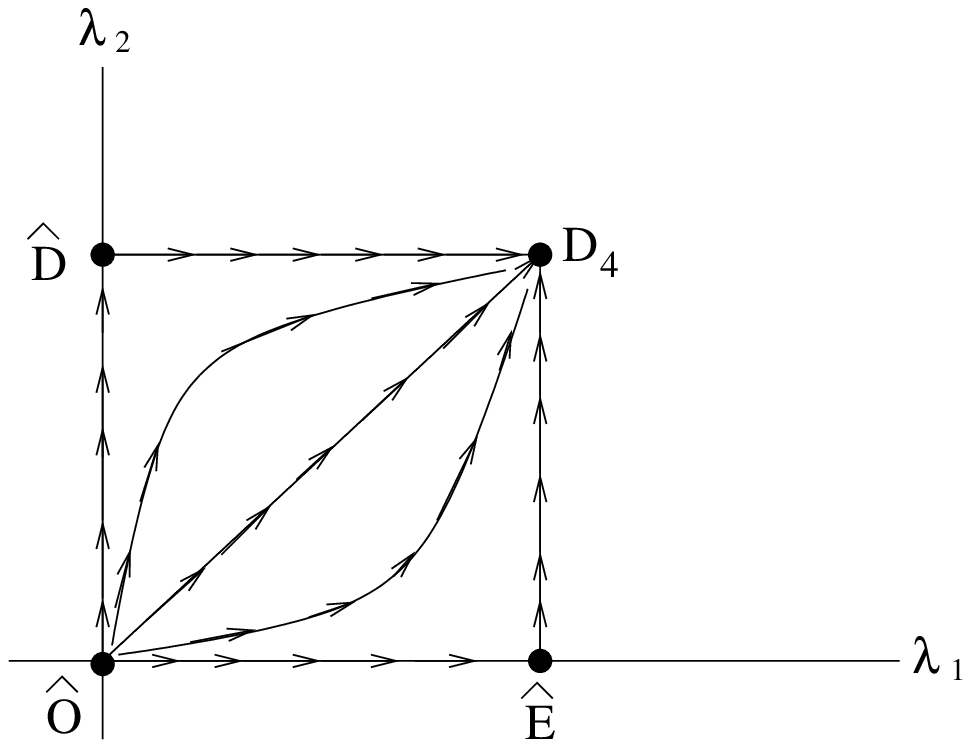}}
\centerline{\ninepoint\sl \baselineskip=8pt {\bf Figure 29:}
{\sl The flow between fixed points in the $\lambda_1$-$\lambda_2$
plane.}}
\bigskip

\appendix{B}{$a$-maximization and the effect of accidental
symmetries}

In \IW, we proved that the unique exact superconformal R-symmetry is
the one that (locally) maximizes the central charge $a$.  The idea
is to write the most general possible R-symmetry, taken to be
anomaly free and respected by relevant superpotential terms, as
$R_t = R_0 + \sum_I s_IJ_I$, where $R_0$ is an arbitrary candidate
R-symmetry, $J_I$ are the various non-R flavor symmetries, and $s_I$
are real parameters.  The superconformal R-symmetry corresponds to
some particular values of the parameters, $\widehat R = R_0+\sum _I
\widehat s_I J_I$, which need to be determined.    The method of
\IW\ for determining $\widehat R$ is to locally maximize the quantity
\eqn\atrial{a_{trial}(s_I) = 3 \Tr R_t^3 - \Tr R_t,}
i.e. find the unique values of $\widehat s_I$ where
\eqn\amx{{\partial a(s) \over \partial s_I} = 0 \quad {\rm and} \quad
{\partial^2 a \over \partial s_I \partial s_J} < 0.}

The above a-maximization proceedure requires knowing the full set of
flavor symmetries $J_I$.  A subset of the flavor symmetry group can be
determined immediately from the classical Lagrangian and anomaly
considerations.  But, if the RG fixed point is at relatively strong
coupling, it is possible that it has additional accidental flavor
symmetries, which are not easily visible in a weak coupling analysis
of the Lagrangian.  One situation where such accidental flavor
symmetries are known to be present is when a chiral gauge invariant
composite operator, $M$, appears otherwise to violate the unitarity
bound $R(M)\geq 2/3$ for all chiral operators
(this follows from the unitarity bound $\Delta (M) \geq 1$).  The
believed
resolution is that $M$ is actually a free field, with $R(M)=2/3$.
There is then an accidental $U(1)_M$ symmetry under which only the
operator $M$ is charged.  $U(1)_M$ mixes with the superconformal
$U(1)_R$ symmetry to make $R(M)=2/3$, without directly affecting the
superconformal R-charge of the other operators.

The a-maximization procedure, however, {\it is} non-trivially affected
by such accidental symmetries \KPS, because the trial $U(1)_R$
symmetry should now include the possibility of mixing with $U(1)_M$:
$R_{t'}=R_t+s_MJ_M$, where $R_t$ is the old trial $U(1)_R$ symmetry;
$R_{t'}$ includes mixing with $U(1)_M$, whereas $R_{t}$ did not.
$J_M$ is the current of the accidental $U(1)_M$ symmetry, which gives
charge $+1$ to the composite gauge invariant operators $M$ (suppose
that there are $dim(M)$ of them), with all other gauge invariant
operators neutral under $J_M$.  The correct a-maximization procedure
is to maximize the combination of 't Hooft anomalies \atrial\ for the
most general trial R-symmetry $R_{t'}$, including mixing with
$U(1)_M$: $a_{t'}(s_I, s_M) =3\Tr R_{t'}^3-\Tr R_{t'}$.  We can use 't
Hooft anomaly matching to account for the difference between $R_{t'}$
and $R_t$, which only comes from the contribution to the 't Hooft
anomalies of the $dim(M)$ free fields $M$:
\eqn\atriala{\eqalign{a_{t'}(s_I, s_M)&=a_{t}(s_I)
+dim(M)\left[3(R_t(M)+s_M-1)^3-
3(R_t(M)-1)^3\right]\cr &-dim(M)\left[(R_t(M)+s_M-1)-(R_t(M)-1)
\right].}}
$a_{t'}(s_I,s_M=0)=3\Tr R_t^3-\Tr R_t\equiv a_t(s_I)$ is the trial
a-function which we would have maximized had there not been the
accidental $U(1)_M$ symmetry.

Upon maximizing $a_t(s_I, s_M)$ in \atriala\ with respect to
$s_M$, the solution $\widehat s_M$ is immediately found to be given by
\eqn\smhat{R_{t'}(M)\equiv R_t(M)+\widehat s_M=2/3,}
which is the expected result that $U(1)_M$ mixes with the superconformal
$U(1)_R$ symmetry precisely so as to make $\widehat R(M)=2/3$.  We can
now maximize $a_{t'}(s_I, \widehat s_M)$ with respect to the other
$s_I$,
to determine their values $\widehat s_I$.  Using \atriala\ and \smhat,
together with
\eqn\rident{3(R(M)-1)^3-(R(M)-1)={2\over 9}-{1\over
9}(2-3R(M))^2(5-3R(M)),}
the quantity which we have to maximize to find the $\widehat s_I$, is
now
\eqn\atrialaa{a_{t'}(s_I,\widehat s_M)=a_{t}(s_I)+
{1\over 9}dim(M)(2-3R_t(M))^2(5-3R_t(M)).}  The presence of the second
term in \atrialaa\ leads to a different maximizing solution $\widehat
s_I$, which non-trivially affects the result for the superconformal
$U(1)_R$ charge of all fields.  It also affects the value of the
central charge $a_{t'}(\widehat s_I, \widehat s_M)$;
maximizing with respect to $s_M$, rather than setting $s_M=0$,
leads to a larger value for the maximal central charge.   The second
term
in \atrialaa\ vanishes when $R(M)=2/3$, so the central charge is
continuous
when $M$ hits the unitarity bound and becomes free.

More generally, as in \KPS, the quantity to maximize will be as in
\atrialaa, but with a sum over every operator $M$ which is a free field.

\appendix{C}{Baryons and the unitarity bound}

In our large $N_c$ limit, baryons can only potentially violate the
unitarity bound $R(B)\geq 2/3$ when $y(x)\equiv R(Q)$ is zero
or negative for some $x$.  This is the case for the $\widehat D$,
$D_{k+2}$, and $E_{6,7,8}$ RG fixed points.  We discuss each
of these now.  If our results had led to baryons violating the unitarity
bound, we would have concluded that those baryons are free fields,
which would
modify our results for the R-charges and central charge $a$.

\subsec{$\widehat D$: no baryons violate the unitarity bound}

For large $x$, $y$ becomes negative and one might worry about baryons
thus having negative R-charge.  The worst-case scenario in terms of
potentially
violating the unitarity bound is a baryon of the form
$$\prod _{i=0}^{[x]-1}(X^iQ)^{N_f}$$
where, to minimize the R-charge, we put in as few $X$'s as possible
and no $Y$'s, bearing in mind that
we need $[x]=N_c / N_f$ different
dressed quarks to antisymmetrize the $N_c$ gauge indices.
  The R-charge of the above baryon is
$$N_cy+N_f\sum _{i=1}^{[x]-1}i{2-2y\over x}=N_cy+N_f\left({1-y\over
x}\right)[x]
([x]-1)\approx N_c,$$
with the terms involving $y$ canceling in the large $x$ limit.  So
no baryons actually ever violate the unitarity bound.

\subsec{$D_{k+2}$: again no baryons violate the unitarity bound.}

The R-charges of the baryons \dkbary\ are given by
\eqn\dkrbary{R(B^{(n_{1,1},\cdots,n_{k,3})}) =
{1 \over k+1} \left [ - N_c (1+x) + \sum_{l=1}^k \sum_{j=1}^3
n_{l,j}(2l+kj)\right ].}  We need to show that this is positive in the
range $k+1 < x < 3k$; for $x<k+1$ the baryons certainly do not violate
unitarity (all R-charges in \rdkis\ are positive), and for $x>3k$
the baryons do not exist (moreover, there is no stable vacuum).

The baryon with smallest R-charge (the worst-case scenario for unitarity
violation) is obtained by minimizing $\sum _{l=1}^k \sum _{j=1}^3
n_{l,j}(2l+kj)$ in \dkrbary, subject to the constraint that $\sum
_{l=1}^k n_{l,j}=N_c$ and $n_{l,j}\leq N_f$.  This is achieved by
taking $n_{l,j}=N_f$ for the $[x]$ different choices of $l$ and $j$
for which $2l+kj$ is as small as possible, with the other $n_{l,j}$
zero.
The ordered list of possibilities for $2l+kj$ is as follows:
\eqn\cljlist{2l+kj=\cases{$k+2, k+4, k+6, \dots , 2k-1, 2k+1$
&taking $l=1\dots \half (k+1)$ with $j=1$\cr
$2k+2,2k+3, 2k+4, \dots, 4k, 4k+1$ &from the others\cr
$4k+3, 4k+5, 4k+7, \dots 5k-2, 5k$ &from $l=\half (k+3)\dots k$ with
$j=3$.}}
We thus have for our worst-case scenerio
\eqn\cljsum{\eqalign{\sum _{l=1}^k\sum _{j=1}^3 n_{l,j}(2l+kj)&=\sum
_{i=1}^{\half (k+1)}N_f(2i+k)+\sum _{i=1}^{min([x]-\half (k+1),2k)}
N_f(2k+1+i)\cr
&+\sum _{i=1}^{[x]-\half (5k+1)}N_f(4k+1+2i),}}
where the last line of \cljsum\ is included only if $[x]>\half (5k+1)$.
Since $[x]>k+1$ in our range of interest, we must use all of the $\half
(k+1)$ terms in the first row of \cljlist and at least some of the
second row, which yield the first two sums in \cljsum.  If $[x]>
\half (5k+1)$ then the second row of \cljlist\ is also used up, and
we need to use the third, giving the last line of \cljsum.  Doing the
sums \cljsum\ and plugging into \dkrbary\ yields the minimal R-charge of
a baryon.  The result is a complicated function of $x$, which starts off
positive at $x=k+1$, initially increases with $x$,
and then monotonically decreases until it reaches 0 at $x=3k$. Thus no
baryons ever violate the unitarity bound for the $\widehat{D}_{k+2}$
theories, at least for $k$ odd.  We expect $k$ even to be qualitatively
similar,
since we can flow from $k$ odd to a lower value of $k'$, which could be
even by adding a superpotential deformation.

\subsec{$E_6$: baryons would violate the unitarity bound for large $x$.}

We now discuss the $E_6$ case; the $E_7$ and $E_8$ cases can be
similarly analyzed.  As we have seen, at the classical level the chiral
ring does not appear to truncate and it is unclear whether or not there
is an upper stability bound on $x$, $x<x^{max}_{E_6}$.   If we ignore
the possible quantum truncation of the chiral ring, along with the
possible
stability bound upper limit on $x$, we can see that there will be
baryons which
hit the unitarity bound, and thus become free fields, for sufficiently
large $x$.  We would then have to correct the $a$ to be
maximized, along the lines of the general discussion in \KPS,
by terms analogous to the additional term in \atrialaa, but for the
baryons.

However, our numerical evidence in sect. 7.1, along with assuming the
a-theorem conjecture is correct, suggests that there is actually
a quantum stability bound requiring $x<x^{max}_{E_6}\approx 13.80$.
And it can be verified that no baryons have yet hit the unitarity bound
if
$x\leq 13.80$.  So, if there is indeed such a stability bound, the
discussion
of this subsection is a moot point.

To see that baryons would hit the unitarity bound for $x$ sufficiently
large (ignoring the possible stability bound mentioned above),
consider the $2^nN_f$ dressed fundamentals
\eqn\dressfe{ X^{r_1}YX^{r_2}Y\dots X^{r_{n-1}}YX^{r_n}YQ_i,}
where each $r_i=1$ or 2.  Forming a baryon from these,
we require $[x]$ types of such dressed quarks, with $n$ ranging from
zero to $q$, given by $\sum _{n=1}^q 2^n=[x]$, i.e. $q\approx \log _2 x$
for large $x$.   The R-charge of such a baryon is
\eqn\revibary{\eqalign{{ R(B)\over N_f} &=xR(Q) + \sum_{n=1}^q n2^n
R(Y) +
\sum_{n=1}^q \sum_{i=0}^n{n \choose i}(i+n)R(X) \cr
&= x \left ( 1-{x \over 6} \right ) + {2 \over 3}\sum_{n=1}^q n2^n +
{1 \over 2}\sum_{n=1}^q \sum_{i=0}^n{n \choose i}(i+n),\cr
&=x \left ( 1-{x \over 6} \right ) +{2\over
3}\left(2+(q-1)2^{q+1}\right)+\half (3+3(q-1)
2^q) . }}  Taking $x$ large for simplicity, this gives
$q\approx \log _2x$, and then \revibary\ yields
\eqn\badbary{{R(B) \over N_f} =
-{x^2 \over 6} + {17x\log x \over 12 \log 2} + x,}
which becomes negative for $x>55$ (a slight overestimate, due to our
approximations).

\appendix{D}{Magnetic Asymptotics}

As at the end of Section 4.1, we can compute the asymptotic value of
$\widetilde{y}$ for the magnetic dual of the $D_{k+2}$ theory. For
large $k$, $\widetilde{x}^{min}_{D_{k+2}}$ also becomes large, and
$\widetilde{y}(\widetilde{x})$ flattens out before the
$\widetilde{X}^{k+1}$ term becomes relevant; it is this asymptotic
value of $\widetilde{y}$ we wish to compute.  The situation is
complicated here by the preponderance of mesons, $3k$ in all. It is
convenient to divide these into 3 groups of $k$, each with different
powers of $Y$.

The central charge $a$ for this theory can then be written as
\eqn\amagdk{\eqalign{{\widetilde{a}_{D_{k+2}}(\widetilde{x},\widetilde{y
})
\over N_f^2}=
{a_{\widehat{D}} (\widetilde{x},\widetilde{y}) \over N_f^2} &+ {1
\over 9}\sum_{l=1}^{p_1}\left [ ( 2 - 3R(N_{l1}))^2
(5-3R(N_{l1}))\right ] +{2 \over 9}(k-2p_1) \cr &+ {1 \over
9}\sum_{l=1}^{p_2}\left [ ( 2 - 3R(N_{l2}))^2 (5-3R(N_{l2}))\right ]
+{2 \over 9}(k-2p_2) \cr & +{1 \over 9}\sum_{l=1}^{p_3}\left [ ( 2 -
3R(N_{l3}))^2 (5-3R(N_{l3}))\right ] +{2 \over 9}(k-2p_3),}} where
\eqn\rnlj{R(N_{lj})=2\widetilde{y}+(l-1){2 - 2\widetilde{y}
\over \widetilde{x}}+
(j-1)\left ( {\widetilde{y}-1\over \widetilde{x}} + 1 \right )} and
$a_{\widehat{D}} (\widetilde{x},\widetilde{y}) / N_f^2$ is the same as
in \axysquared.  The sums run over the Legendre-transform partners
$N_{lj}$ of the interacting mesons $M_{k+1-l,3-j}$, and we have split
the sum over the remaining free mesons into three parts; the total
number of free mesons is $3k -p_1 - p_2 - p_3$.

We would now like to approximate these sums by integrals. However, it
is not immediately clear which sums we should include. We know that
the meson $M_{k3}$ is interacting for all $\widetilde x > 1$, so we
should definitely include the $p_1$ sum. But it is not obvious whether
or not to include the $p_2$ and $p_3$ sums.  We use a
self-consistency check to give us the answer:
The upper limit of the
sums is the value at which the meson with the smallest R-charge
becomes interacting,
\eqn\rmint{2\widetilde{y}+(p_j -1)\left ( {2 - 2\widetilde{y}
\over \widetilde{x}} \right )
+(j-1) \left ( {\widetilde{y}-1\over \widetilde{x}} + 1 \right ) = {4
\over 3}.}  (Note: This looks like it describes when the
Legendre-transform partner with the largest possible R-charge $N_{lj}$
becomes free, except for the 4/3 on the RHS.)  Since we expect
$\widetilde{y}$ to approach a constant, we can easily solve this for
$p_j$ in the large $\widetilde{x}$ limit:
\eqn\pnsol{p_j = \left ( {4 \over 3} -j+1-2 \widetilde{y} \right )
\left ( {\widetilde{x} \over 2 - 2\widetilde{y}} \right).}

To approximate these sums by integrals, we can integrate over the
variable
\eqn\ti{t_j = 2\widetilde{y}+(l -1)\left ( {2 - 2\widetilde{y} \over
\widetilde{x}} \right )
+(j-1) \left ( {\widetilde{y}-1\over \widetilde{x}} + 1 \right );} the
limits of integration on the $n$th sum will be from $2\widetilde{y}
+n-1$ to $4/3$.  Performing these integrals and substituing the answer
back into \amagdk\ gives an expression for $a$ which may be maximized
as a function of $\widetilde{y}$.

The self-consistency argument is easily done: First, assume that only
the $p_1$ sum contributes.  Including its contribution to
$\widetilde{a}$ and maximizing yields a number for $\widetilde{y}$
which may then be substituted into \pnsol. Doing so yields positive
values for both $p_1$ and $p_2$, which is inconsistent with our
assumption that only the first sum contributes. One may then perform
this procedure for any combination of the sums; the only
self-consistent answer turns out to be that one must include both the
$p_1$ and $p_2$ sum, but not the $p_3$ sum.  Maximizing
$\widetilde{a}$ with these two sums (and, of course, their
accompanying sums over the free mesons) gives the asymptotic value for
$\widetilde{y}$
\eqn\asympy{\widetilde{y}_{asymp} \approx -0.1038.}
One may then easily compute the asymptotic value of the central charge
\eqn\asympa{{\widetilde{a}_{D_{k+2}}(\widetilde{x}) \over N_f^2}
\approx 13.1186\widetilde{x}+{6k \over 9}}
and also where the $\widetilde{X}^{k+1}$ term becomes relevant,
\eqn\asympx{\widetilde{x}_{D_{k+2}}^{min} \approx 1.1038 k.}

\listrefs
\end